\documentclass[12pt]{iopart}

\usepackage{iopams}  
\expandafter\let\csname equation*\endcsname\relax
\expandafter\let\csname endequation*\endcsname\relax
\usepackage{amsmath,amssymb}
\usepackage{physics}
\usepackage{cite}
\usepackage[pdftex]{graphicx} % for arXiv
\usepackage{bm}% bold math
\usepackage[T1]{fontenc}
\usepackage[pdftex]{color, hyperref}% for arXiv add hypertext capabilities
\hypersetup{
colorlinks = true,
linkcolor = blue,
citecolor = blue
}
\usepackage{here} %図Here位置強制
\usepackage{color}

\begin{document}

\title[Scale locality of information flow in shell models of turbulence]{Scale locality of information flow in shell models of turbulence}

\author{Tomohiro Tanogami}

\address{Department of Earth and Space Science, The University of Osaka, Osaka 560-0043, Japan}
\ead{tanogami@ess.sci.osaka-u.ac.jp}
\vspace{10pt}
\begin{indented}
\item[]\today
\end{indented}

\begin{abstract}
%%%%%% Introduction
Turbulent fluctuations exhibit universal scaling laws that are independent of large-scale statistics.
It is often explained that such universality is caused by the loss of ``information'' about large-scale statistics during the cascade process.
%%%%%% Details of motivation/previous studies
In our previous study [T.~Tanogami and R.~Araki, Phys.~Rev.~Research \textbf{6}, 013090 (2024)], we applied information thermodynamics to turbulence and proved that information of large-scale turbulent fluctuations is propagated to small scales.
%%%%%% Research question
As a first step toward understanding how universality emerges at small scales under the influence of the information flow from large scales, here we investigate the scale locality of the information flow for shell models.
%%%%%% Result
First, we analytically show that the information flow can be decomposed into scale-local and scale-nonlocal parts.
Then, by assuming the Kolmogorov hypothesis for the Kolmogorov multiplier, we prove that the scale-nonlocal part can be ignored compared to the scale-local part.
This result implies that the information transfer from large to small scales occurs mainly through scale-local interactions, which is consistent with the scale locality of the energy cascade.
%%%%%% Details of the result
%%%%%% Significance/impact
%%%%%% Implications
\end{abstract}

%\maketitle
% \tableofcontents

\section{Introduction}
%%%%%% Background: Universal statistics of turbulent fluctuations
In fully developed three-dimensional turbulence, turbulent fluctuations exhibit universal scaling laws that are independent of large-scale statistics~\cite{Frisch,davidson2015turbulence,Eyink_lecture}.
For example, the energy spectrum follows the Kolmogorov spectrum $\propto k^{-5/3}$ in the inertial range $k_f\ll k\ll k_\nu$, where $k_f$ and $k_\nu$ denote the energy injection and dissipation scales, respectively.
In this intermediate-scale range, both forcing and viscous effects are negligible, and the energy is transferred conservatively from large to small scales.
This energy transfer across scales is called the energy cascade and is considered to be the generating mechanism of the universal scaling laws.
In particular, it is often explained that the universal scaling laws are caused by the loss of ``information'' about large-scale statistics during the energy cascade process~\cite{davidson2015turbulence,Eyink_Sreenivasan,Eyink_lecture}.

%%%%%% Background: Information flow in turbulence
Somewhat contrary to this intuitive picture, our previous study proved that information of turbulent fluctuations at large scales is transferred to small scales in the inertial range~\cite{tanogami2024information,tanogami2025amplify}.
Here, the information is quantified by the mutual information that measures the strength of the correlation between large- and small-scale turbulent fluctuations (a precise definition is provided in Sec.~\ref{sec: Scale-to-scale information flow})~\cite{cover1999elements}.
Because this result is a direct consequence of the second law of information thermodynamics, which is a thermodynamic framework for information flow between interacting subsystems~\cite{peliti2021stochastic,shiraishi2023introduction,parrondo2015thermodynamics,horowitz2014thermodynamics,ehrich2023energy}, the result is exact and universal, independent of the details of the flow under consideration.
In particular, the result holds for various thermodynamically consistent turbulence models that exhibit the energy cascade, such as the Navier--Stokes equation and shell models with thermal noise.

\begin{figure}[t]
\centering
\includegraphics[width=10cm]{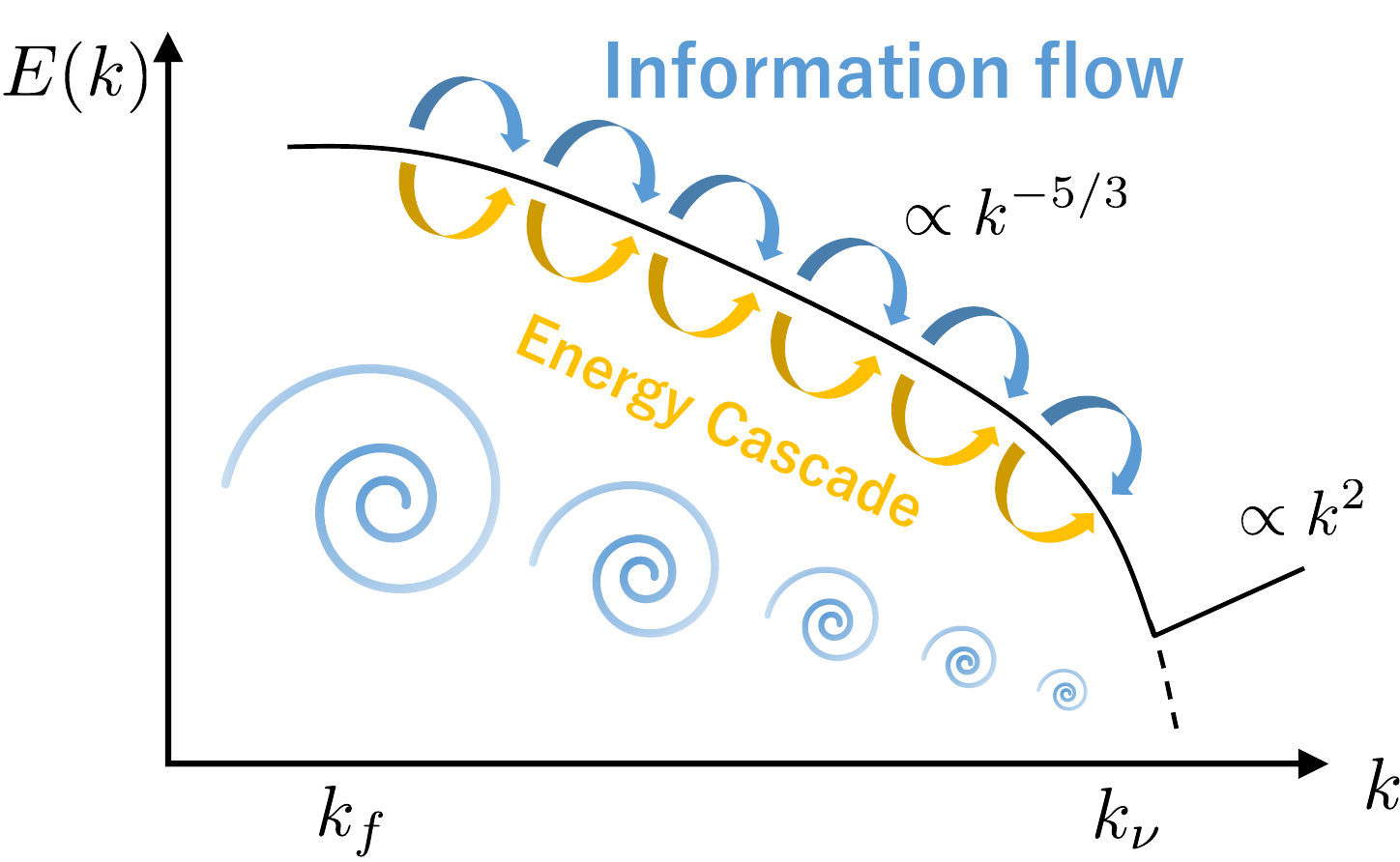}
\caption{Schematic of the scale locality of the information flow in turbulence.
The solid black line represents the energy spectrum, which follows the Kolmogorov spectrum $E(k)\propto k^{-5/3}$ in the inertial range $k_f\ll k\ll k_\nu$.
Here, $k_f$ and $k_\nu$ denote the energy injection and dissipation scales, respectively.
The blue and yellow arrows indicate the direction and scale locality of the scale-to-scale information flow $\dot{\mathcal{I}}_K$ and the energy cascade, respectively.
In the dissipation range, the spectrum decays exponentially in the absence of thermal noise (dashed line), or it exhibits equipartition of energy as $\propto k^2$ when thermal noise is present (solid line)~\cite{bandak2022dissipation,bell2022thermal,mcmullen2022navier}.
}
\label{fig:schematic_scale_locality}
\end{figure}

%%%%%% Research question & Approach
The fact that information of turbulent fluctuations is propagated from large to small scales suggests that turbulent fluctuations at small scales are influenced by those at large scales.
To put it another way, our previous result raises a new question about how universal scaling laws of turbulent fluctuations emerge at small scales under the influence of the information flow from large scales.
As a first step toward solving this problem, here we investigate the scale locality of the information flow for shell models, which are simplified models of the Navier--Stokes equation.
The scale locality means that the information of turbulent fluctuations is not directly transferred from the largest to the smallest scales but mainly through scale-local interactions (see Fig.~\ref{fig:schematic_scale_locality}).
Here, we note that the scale locality of the \textit{energy} cascade has been studied over the past few decades because the scale locality has been considered to be the fundamental mechanism behind the universal statistics at small scales~\cite{domaradzki1990local,eyink2005locality,eyink2009localness,aluie2009localness,cardesa2017turbulent,goto2017hierarchy,johnson2020energy,johnson2021role}.
We expect that the scale locality of the \textit{information} transfer is similarly or even more essential to the generation mechanism of the universal statistics.
Indeed, if the information transfer is scale-nonlocal, the small scales may strongly depend on the large-scale statistics, implying non-universality at small scales.
In contrast, if the information flow satisfies the scale locality, then we can conjecture that the universal scaling laws of turbulent fluctuations at small scales result from the accumulation of some kind of errors during the scale-local transmission process.
% From the analogy with the picture associated with the scale locality of the energy cascade, we conjecture that the universal scaling laws are related to the scale locality of the information flow
% Although the scale locality of the \textit{energy} cascade has been studied over the past few decades~\cite{domaradzki1990local,eyink2005locality,eyink2009localness,aluie2009localness,cardesa2017turbulent,goto2017hierarchy,johnson2020energy,johnson2021role}, the scale locality of the \textit{information} transfer may be more essential to the generation mechanism of the universal scaling laws.

%%%%%% Summary of results
In this paper, we prove that the information flow in shell models satisfies the scale locality in the inertial range.
More specifically, we first analytically show that the information flow can be decomposed into scale-local and scale-nonlocal parts.
Using this decomposition and assuming the Kolmogorov hypothesis for the Kolmogorov multiplier~\cite{kolmogorov1962refinement}, we then show that the scale-nonlocal part can be ignored compared to the scale-local part.
That is, the information transfer from large to small scales occurs mainly through scale-local interactions, which is consistent with the scale locality of the energy cascade.
Although our results are restricted to shell models, we conjecture that similar results can be obtained for other turbulence models, including the Navier--Stokes equation, because the Kolmogorov hypothesis agrees well with experimental and direct numerical simulation results~\cite{chen2003kolmogorov,benzi1998multiscale,pedrizzetti1996self,chhabra1992scale}.

%%%%%% Organization of the paper
The rest of this paper is organized as follows.
In Sec.~\ref{Setup}, we introduce the Sabra shell model with thermal noise and its corresponding Fokker--Planck equation.
In Sec.~\ref{Information-theoretic quantities}, we first introduce mutual information and information flow, the key information-theoretic quantities in this study, in a general setup.
We then define the \textit{scale-to-scale information flow} for the shell model and discuss its relevant properties, including our previous results obtained in Ref.~\cite{tanogami2024information}.
We also introduce the decomposition of the scale-to-scale information flow into reversible and irreversible parts, which is used to prove the scale locality.
In Sec.~\ref{Scale locality of the information flow}, we discuss the scale locality of the information flow.
We first show that the scale-to-scale information flow can be decomposed into scale-local and scale-nonlocal parts, and then prove that the scale-nonlocal part can be ignored compared to the scale-local part.
% Because the proof is rather long and technical, only an outline of the proof is provided in this section.
Finally, in Sec.~\ref{Concluding remarks}, we summarize our main results with some remarks.
The Appendices contain the details of the derivation.

\section{Setup\label{Setup}}
In this study, we consider the Sabra shell model, although the same results can also be obtained for other shell models with scale-local nonlinear interactions, such as the GOY model~\cite{gledzer1973system,ohkitani1989temporal}.
We also explicitly consider thermal fluctuations, an effect intrinsic to molecular fluids that arises from microscopic random motions.
In our previous studies~\cite{tanogami2024information,tanogami2025amplify}, thermal fluctuations were taken into account to make the model thermodynamically consistent and suitable for analysis using information thermodynamics.
Although thermal fluctuations have been considered to have only a negligible effect on turbulence dynamics, recent studies have reported that thermal noise significantly affects statistics in the dissipation range and predictability of turbulence~\cite{bandak2022dissipation,bell2022thermal,komatsu2014glimpse,mcmullen2022navier,ruelle1979microscopic,lorenz1969predictability,bandak2024spontaneous,palmer2024real,Srivastava2025_molecular}.
Thus, incorporating thermal fluctuations is a suitable approach for our purpose, from both thermodynamic and turbulence dynamics perspectives.
% Here, to clarify the connection with our previous study~\cite{tanogami2024information,tanogami2025amplify}, we explicitly consider thermal fluctuations, the effect intrinsic to molecular fluids.
% , which included thermal noise to make the model thermodynamically consistent~\cite{peliti2021stochastic}.
% Recent studies have reported that thermal noise significantly affects statistics in the dissipation range and predictability of turbulence~\cite{bandak2022dissipation,bell2022thermal,komatsu2014glimpse,mcmullen2022navier,ruelle1979microscopic,lorenz1969predictability,bandak2024spontaneous,palmer2024real,Srivastava2025_molecular}.
% We emphasize, however, that the effect of thermal noise is not essential to the results described in this paper; the scale locality of the information flow can also be proved for the deterministic case.
% For more details on this point, see Sec.~\ref{Decomposition of the information flow into reversible and irreversible parts}.

We now introduce the fluctuating Sabra shell model and its corresponding Fokker--Planck equation.
Let $u_n(t)\in\mathbb{C}$ be the ``velocity'' at time $t$ in a shell of wavenumber $k_n=k_02^n$ ($n=0,1,\cdots,N$).
We denote by $u^*_n(t)$ the complex conjugate of $u_n(t)$.
The time evolution of the complex shell variables $u:=\{u_n\}$ is given by the following Langevin equation~\cite{l1998improved,bandak2021thermal,bandak2022dissipation}:
\begin{align}
\partial_t{u}_n=B_n(u,u^*)-\nu k^2_nu_n+\sqrt{\dfrac{2\nu k^2_nk_{\mathrm{B}}T}{\rho}}\xi_n+f_n,
\label{sabra shell model}
\end{align}
where $B_n(u,u^*)$ denotes the scale-local nonlinear interactions defined by
\begin{align}
B_n(u,u^*)&:=i\biggl(k_{n+1}u_{n+2}u^*_{n+1}-\dfrac{1}{2}k_nu_{n+1}u^*_{n-1}+\dfrac{1}{2}k_{n-1}u_{n-1}u_{n-2}\biggr)
\end{align}
with $u_{-1}=u_{-2}=u_{N+1}=u_{N+2}=0$, $\nu$ represents ``(bare) kinematic viscosity'', $f_n\in\mathbb{C}$ denotes the external body force that acts only at large scales, i.e., $f_n=0$ for $n>n_f$.
For simplicity, we assume that $f_n$ is independent of time.
The third term on the right-hand side of Eq.(\ref{sabra shell model}) denotes thermal noise, where $\xi_n\in\mathbb{C}$ is the zero-mean white Gaussian noise that satisfies $\langle\xi_n(t)\xi^*_{n'}(t')\rangle=2\delta_{nn'}\delta(t-t')$, $T$ denotes the absolute temperature, $k_{\mathrm{B}}$ the Boltzmann constant, and $\rho$ the mass ``density''.
Note that $\rho$ has units of mass in the shell model.
The strength of the noise term satisfies the fluctuation-dissipation relation of the second kind~\cite{maes2021local,peliti2021stochastic}.
Note that the fluctuating Sabra shell model~(\ref{sabra shell model}) can be regarded as a simplified model of the fluctuating Navier--Stokes equation~\cite{landau1959fluid,de2006hydrodynamic,bandak2022dissipation}.

Let $p_t(u,u^*)$ be the probability distribution of state $(u,u^*)$ at time $t$.
The time evolution of $p_t(u,u^*)$ is governed by the following Fokker--Planck equation~\cite{risken1996fokker}, which is equivalent to the Langevin equation~(\ref{sabra shell model}):
\begin{align}
\partial_tp_t(u,u^*)=\sum^N_{n=0}\left[-\dfrac{\partial}{\partial u_n}J_n(u,u^*)-\dfrac{\partial}{\partial u^*_n}J^*_n(u,u^*)\right],
% &=\sum^N_{n=0}\left[-\dfrac{\partial}{\partial u_n}\left(A_n(u,u^*)p_t(u,u^*)\right)-\dfrac{\partial}{\partial u^*_n}\left(A^*_n(u,u^*)p_t(u,u^*)\right)+\dfrac{4\nu k^2_nk_{\mathrm{B}}T}{\rho}\dfrac{\partial^2}{\partial u_n\partial u^*_n}p_t(u,u^*)\right]\notag\\
\label{FP-sabra}
\end{align}
where $J_n(u,u^*)$ denotes the probability current associated with the shell variable $u_n$:
\begin{align}
J_n(u,u^*)&:=\left(B_n(u,u^*)-\nu k^2_nu_n+f_n\right)p_t(u,u^*)-\dfrac{2\nu k^2_nk_{\mathrm{B}}T}{\rho}\dfrac{\partial}{\partial u^*_n}p_t(u,u^*).
\label{probability current}
\end{align}

When $f_n=0$ for all $n$, the system relaxes to a thermal equilibrium state, which is described by the Gibbs distribution:
\begin{align}
p_{\mathrm{eq}}(u,u^*)=\dfrac{1}{Z}\exp\left(-\dfrac{1}{k_{\mathrm{B}}T}\sum^N_{n=0}\dfrac{1}{2}\rho|u_n|^2\right),
\label{Gibbs distribution}
\end{align}
where $Z$ is a normalization constant.
In contrast, when $f_n\neq0$ for $n\le n_f$, the system approaches to an nonequilibrium steady state, which is described by a steady-state distribution $p_{\mathrm{ss}}(u,u^*)$ that is distinct from $p_{\mathrm{eq}}(u,u^*)$.
In particular, if there is a scale separation between the energy injection scale $k_f:=k_{n_f}$ and the energy dissipation scale $k_\nu:=\nu^{-3/4}\varepsilon^{1/4}$, where $\varepsilon:=\sum^N_{n=0}\nu k^2_n\langle|u_n|^2\rangle$ denotes the energy dissipation rate, the system generally exhibits the energy cascade and universal scaling laws in the inertial range $k_f\ll k_n\ll k_\nu$~\cite{bohr1998dynamical,l1998improved,bandak2022dissipation,biferale2003shell}.
% In this model, the energy injection and dissipation scales are defined as $k_f:=k_{n_f}$ and $k_\nu:=\nu^{-3/4}\varepsilon^{1/4}$, respectively, where $\varepsilon$ denotes the energy dissipation rate defined by $\varepsilon:=\sum^N_{n=0}\nu k^2_n\langle|u_n|^2\rangle$.

\section{Information-theoretic quantities\label{Information-theoretic quantities}}
In this section, we introduce important information-theoretic quantities for this study.
In Sec.~\ref{sec: Mutual information and information flow}, we define mutual information and information flow in a general setup.
Then, in Sec.~\ref{sec: Scale-to-scale information flow}, we introduce \textit{scale-to-scale information flow} for turbulence and describe its relevant properties.
We also briefly review the information-thermodynamic bound on the scale-to-scale information flow, which was obtained in our previous paper~\cite{tanogami2024information}.
In Sec.~\ref{Decomposition of the information flow into reversible and irreversible parts}, we introduce the decomposition of the scale-to-scale information flow into reversible and irreversible parts.
This decomposition will be used to prove the scale locality in the next section.

\subsection{Mutual information and information flow\label{sec: Mutual information and information flow}}
While several previous studies have attempted to quantify statistical causality in turbulence using other information-theoretic quantities, such as (net) \textit{transfer entropy}~\cite{schreiber2000measuring,materassi2014information} or \textit{information flux}~\cite{PhysRevResearch.4.023195,araki2024forgetfulness}, these quantities can be nonzero even in an equilibrium state where all probability currents vanish~\cite{chetrite2019information}.
For other shortcomings, see, e.g., Refs.~\cite{smirnov2013spurious,james2016information}.
% Although many other information-theoretic quantities have been applied to turbulence, such as \textit{transfer entropy}~\cite{schreiber2000measuring,materassi2014information,PhysRevResearch.4.023195,araki2024forgetfulness}, \textit{time-delayed mutual information}~\cite{cerbus2016information,goldburg2016turbulence}, \textit{entropy rate}~\cite{cerbus2013information,goldburg2016turbulence,granero2016scaling}, \textit{Kullback--Leibler divergence}~\cite{granero2018kullback}, the information flow is appropriate for our argument because it can quantify causal relationships and appears in the second law of information thermodynamics~\cite{horowitz2014thermodynamics,ehrich2023energy}.
Here, we instead employ \textit{information flow} (also called \textit{learning rate})~\cite{allahverdyan2009thermodynamic,hartich2014stochastic,barato2014efficiency,horowitz2015multipartite,hartich2016sensory,matsumoto2018role}, which is defined through the decomposition of the time derivative of mutual information.
Because the information flow becomes zero in an equilibrium state where all probability currents are absent, it is expected to properly describe the causal influences in nonequilibrium states. 
Below, we introduce the mutual information and information flow in a general setup.

\subsubsection{Mutual information}
Let $X$ and $Y$ be two continuous random variables, and $p(x,y)$ be their joint probability density.
The marginal distributions are given by $p^X(x):=\int dyp(x,y)$ and $p^Y(y):=\int dxp(x,y)$.
The mutual information is defined by~\cite{cover1999elements}
\begin{align}
I[X\colon\!Y]:=\int dxdyp(x,y)\ln\dfrac{p(x,y)}{p^X(x)p^Y(y)}.
\end{align}
Here, we use the notation $I[X\colon\!Y]$ to explicitly indicate the relevant random variables $X$ and $Y$, although this is not a function of these variables.
The mutual information can be expressed in terms of the Kullback--Leibler divergence $D_{\mathrm{KL}}(p\|q):=\int d\omega p(\omega)\ln(p(\omega)/q(\omega))$, which quantifies the ``distance'' between the two probability densities $p$ and $q$~\cite{cover1999elements}:
\begin{align}
I[X\colon\!Y]=D_{\mathrm{KL}}\left(p\|p^Xp^Y\right)\ge0.
\end{align}
Hence, $I[X\colon\!Y]$ can be interpreted as a ``distance'' between the joint probability distribution $p(x,y)$ and the product of the marginal distributions $p^X(x)p^Y(y)$.
In other words, the mutual information quantifies mutual dependence between $X$ and $Y$.
The mutual information is nonnegative and is equal to zero if and only if $X$ and $Y$ are statistically independent.
If there is an additional random variable $Z$, we can also define the \textit{conditional mutual information} $I[X\colon\!Y|Z]$ as~\cite{cover1999elements}
\begin{align}
I[X\colon\!Y|Z]:=\int dxdydzp(x,y,z)\ln\dfrac{p(x,y|z)}{p^X(x|z)p^Y(y|z)},
\end{align}
where $p(x,y|z)=p(x,y,z)/p(z)$ denotes the conditional probability density, and $p^X(x|z)=\int dyp(x,y|z)$ and $p^Y(y|z)=\int dxp(x,y|z)$ denote the marginal conditional distributions.
Conditional mutual information quantifies the strength of the direct correlation between two random variables, $X$ and $Y$, without involving $Z$.

\subsubsection{Information flow}
% Next, we define the information flow.
Note that the mutual information is symmetric between the two variables; therefore, it cannot quantify the directional flow of information from one variable to the other.
The directional flow of information can be quantified in terms of information flow, which is defined through the decomposition of the time derivative of the mutual information.
We denote by $X_t$ and $Y_t$ the random variables at time $t$.
% The directional flow of information can be quantified in terms of information flow, which is also called the \textit{learning rate}~\cite{allahverdyan2009thermodynamic,hartich2014stochastic,barato2014efficiency,horowitz2015multipartite,hartich2016sensory,matsumoto2018role}.
Then, the information flow from $Y$ to $X$ at time $t$ is defined by
\begin{align}
\dot{I}^X[X\colon\!Y]:=\lim_{\Delta t\rightarrow0^+}\dfrac{I[X_{t+\Delta t}\colon\!Y_t]-I[X_t\colon\!Y_t]}{\Delta t}.\label{IF definition X}
\end{align}
The information flow $\dot{I}^X[X\colon\!Y]$ denotes the rate at which $X$ gains information about $Y$.
Similarly, the information flow from $X$ to $Y$ is defined by
\begin{align}
\dot{I}^Y[X\colon\!Y]:=\lim_{\Delta t\rightarrow0^+}\dfrac{I[X_t\colon\!Y_{t+\Delta t}]-I[X_t\colon\!Y_t]}{\Delta t}.\label{IF definition Y}
\end{align}
If the two stochastic processes $X_t$ and $Y_t$ are \textit{bipartite}, i.e., if the transition probability $p(x_{t+dt},y_{t+dt}|x_t,y_t)$ satisfies
\begin{align}
p(x_{t+dt},y_{t+dt}|x_t,y_t)=p(x_{t+dt}|x_t,y_t)p(y_{t+dt}|x_t,y_t)+O(dt^2),
\end{align}
then the sum of $\dot{I}^X[X\colon\!Y]$ and $\dot{I}^Y[X\colon\!Y]$ equals the time derivative of the mutual information~\cite{chetrite2019information}:
\begin{align}
\dfrac{d}{dt}I[X\colon\!Y]=\dot{I}^X[X\colon\!Y]+\dot{I}^Y[X\colon\!Y].
\label{dtI_Ix_Iy}
\end{align}
In the case of diffusion processes, the bipartite condition is satisfied if the noises acting on $X$ and $Y$ are uncorrelated.
Note that in the steady state, $\dot{I}^X[X\colon\!Y]=-\dot{I}^Y[X\colon\!Y]$ because $d_tI[X\colon\!Y]=0$.

The sign of the information flow indicates the direction in which the information is transmitted.
If $\dot{I}^X[X\colon\!Y]>0$, then it means that the dynamical evolution of $X$ increases the correlation between $X$ and $Y$.
More intuitively, $X$ is ``learning'' about $Y$ through its dynamical evolution, and information is transferred from $Y$ to $X$.
In contrast, $\dot{I}^X[X\colon\!Y]<0$ means that the dynamical evolution of $X$ decreases the correlation between $X$ and $Y$.
In this case, $X$ is destroying the correlation with $Y$ through its dynamical evolution.

If there is an additional random variable $Z_t$, we can also define the \textit{conditional information flow} through the time derivative of the conditional mutual information $I[X\colon\!Y|Z]$:
\begin{align}
\dot{I}^X[X\colon\!Y|Z]&:=\lim_{\Delta t\rightarrow0^+}\dfrac{I[X_{t+\Delta t}\colon\!Y_t|Z_t]-I[X_t\colon\!Y_t|Z_t]}{\Delta t},\label{conditional IF definition X}\\
\dot{I}^Y[X\colon\!Y|Z]&:=\lim_{\Delta t\rightarrow0^+}\dfrac{I[X_t\colon\!Y_{t+\Delta t}|Z_t]-I[X_t\colon\!Y_t|Z_t]}{\Delta t}.\label{conditional IF definition Y}
\end{align}
These quantities denote the direct transfer of information between $X$ and $Y$ without involving $Z$~\cite{horowitz2015multipartite}.
Note that, in general, 
\begin{align}
\dot{I}^X[X\colon\!Y|Z]\neq-\dot{I}^Y[X\colon\!Y|Z]
\label{conditional IF neq}
\end{align}
even in the steady state.
In other words, the sum of these two conditional information flows is not equal to the time derivative of the conditional mutual information because there is a contribution from the dynamical evolution of $Z$:
\begin{align}
\dfrac{d}{dt}I[X\colon\!Y|Z]\neq\dot{I}^X[X\colon\!Y|Z]+\dot{I}^Y[X\colon\!Y|Z].
\end{align}

If the time evolution of the joint probability distribution for $X_t$ and $Y_t$ is governed by the Fokker--Planck equation,
\begin{align}
\partial_tp_t(x,y)=-\partial_xJ^X_t(x,y)-\partial_yJ^Y_t(x,y),
\end{align}
where $J^\alpha_t(x,y)$ denotes the probability current associated with $\alpha=X,Y$, then the information flow can be expressed in terms of the probability current as
\begin{align}
\dot{I}^X[X\colon\!Y]&=\int dxdyJ^X_t(x,y)\partial_x\ln\dfrac{p_t(x,y)}{p^X_t(x)p^Y_t(y)},\label{IF expression J_X}\\
\dot{I}^Y[X\colon\!Y]&=\int dxdyJ^Y_t(x,y)\partial_y\ln\dfrac{p_t(x,y)}{p^X_t(x)p^Y_t(y)}.\label{IF expression J_Y}
\end{align}
The derivation of these expressions is provided in~\ref{Expression for information flow in terms of probability currents}.
From these expressions, we can see that the information flow becomes zero in a steady state when the two probability currents are zero, which corresponds to an equilibrium state.
Similarly, if the time evolution of the joint probability distribution $p_t(x,y,z)$ is governed by the Fokker--Planck equation of the form $\partial_tp_t(x,y,z)=-\sum_\alpha \partial_\alpha J^\alpha_t(x,y,z)$ ($\alpha=X,Y,Z$), the conditional information flow can be expressed as
\begin{align}
\dot{I}^X[X\colon\!Y|Z]&=\int dxdyJ^X_t(x,y,z)\partial_x\ln\dfrac{p_t(x,y|z)}{p^X_t(x|z)p^Y_t(y|z)},\label{conditional IF expression J_X}\\
\dot{I}^Y[X\colon\!Y|Z]&=\int dxdyJ^Y_t(x,y,z)\partial_y\ln\dfrac{p_t(x,y|z)}{p^X_t(x|z)p^Y_t(y|z)}.\label{conditional IF expression J_Y}
\end{align}
These expressions are derived in essentially the same way as Eqs.~(\ref{IF expression J_X}) and (\ref{IF expression J_Y}).

\subsection{Scale-to-scale information flow\label{sec: Scale-to-scale information flow}}
\begin{figure}[t]
\centering
\includegraphics[width=10cm]{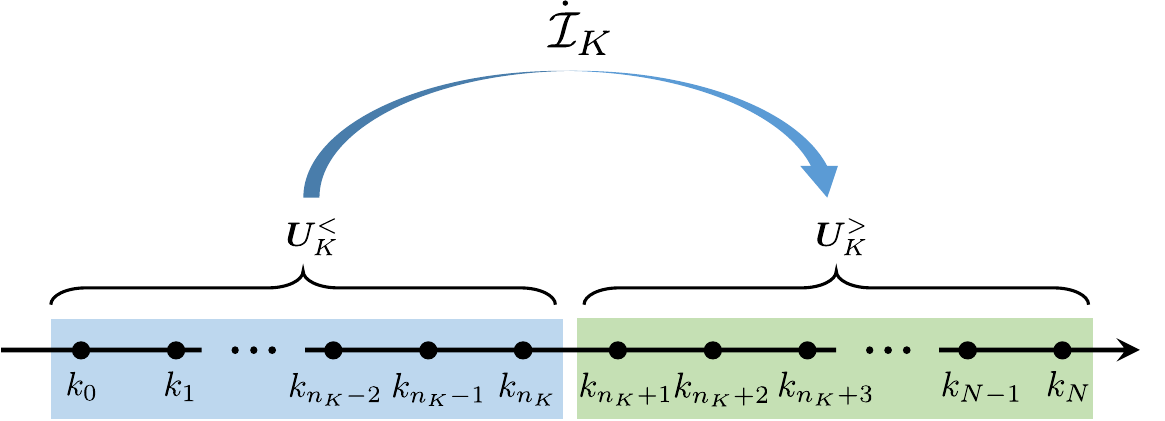}
\caption{Schematic of the information flow across scales.
The total shell variables $\{u,u^*\}:=\{u_n,u^*_n\}^N_{n=0}$ are divided into the large-scale modes ${\bm U}^<_K$ and small-scale modes ${\bm U}^>_K$ at an arbitrary intermediate scale $K:=k_{n_K}$ with $n_K\in\{0,\cdots,N\}$.
If the information flow $\dot{\mathcal{I}}_K$ is positive, then it means that the small-scale modes ${\bm U}^>_K$ are gaining information about the large-scale modes ${\bm U}^<_K$, as shown by the blue arrow.}
\label{fig:Fourier_modes_division}
\end{figure}

Here, we introduce the \textit{scale-to-scale information flow} for turbulence, which was first introduced in our previous paper to quantify information flow across scales~\cite{tanogami2024information,tanogami2025amplify}.
We first divide the total shell variables $\{u,u^*\}$ into two parts at an arbitrary intermediate scale $K:=k_{n_K}$ with $n_K\in\{0,\cdots,N\}$:
\begin{align}
\{u,u^*\}={\bm U}^<_K\cup{\bm U}^>_K,
\end{align}
where ${\bm U}^<_K:=\{u_n,u^*_n\mid0\le n\le n_K\}$ and ${\bm U}^>_K:=\{u_n,u^*_n\mid n_K< n\le N\}$ denote the large-scale and small-scale modes, respectively (see Fig.~\ref{fig:Fourier_modes_division}).

The strength of the correlation between the large-scale modes ${\bm U}^<_K$ and small-scale modes ${\bm U}^>_K$ at time $t$ is quantified by the mutual information defined by
\begin{align}
% I[{\bm U}^<_K\colon\!{\bm U}^>_K]:=\int d{\bm U}^<_Kd{\bm U}^>_Kp_t({\bm U}^<_K,{\bm U}^>_K)\ln\dfrac{p_t({\bm U}^<_K,{\bm U}^>_K)}{p_t({\bm U}^<_K)p_t({\bm U}^>_K)},
I[{\bm U}^<_K(t)\colon\!{\bm U}^>_K(t)]:=\left\langle\ln\dfrac{p_t({\bm U}^<_K,{\bm U}^>_K)}{p_t({\bm U}^<_K)p_t({\bm U}^>_K)}\right\rangle,
\label{def: MI}
\end{align}
where $\langle\cdot\rangle$ denotes the average with respect to the joint probability distribution $p_t({\bm U}^<_K,{\bm U}^>_K)$, and $p_t({\bm U}^<_K)$ and $p_t({\bm U}^>_K)$ are the marginal distributions for the large-scale and small-scale modes, respectively.
Note that the joint probability distribution $p_t({\bm U}^<_K,{\bm U}^>_K)$ is nothing but the probability density for the total shell variables $p_t(u,u^*)$ governed by the Fokker--Planck equation (\ref{FP-sabra}).
At the equilibrium state described by the Gibbs distribution (\ref{Gibbs distribution}), the mutual information becomes zero because the large-scale and small-scale modes are independent.
In contrast, at the nonequilibrium state characterized by the energy cascade, the mutual information is positive, reflecting mutual dependence across scales~\cite{tanogami2024information}.

Next, we define the \textit{scale-to-scale information flow} $\dot{I}^<_K[{\bm U}^<_K\colon\!{\bm U}^>_K]$ at scale $K$ as the information flow from the small-scale modes ${\bm U}^>_K$ to the large-scale modes ${\bm U}^<_K$:
\begin{align}
\dot{I}^<_K[{\bm U}^<_K\colon\!{\bm U}^>_K]:=\lim_{\Delta t\rightarrow0}\dfrac{I[{\bm U}^<_K(t+\Delta t)\colon\!{\bm U}^>_K(t)]-I[{\bm U}^<_K(t)\colon\!{\bm U}^>_K(t)]}{\Delta t}.
% &=\sum^{n_K}_{n=0}\int dudu^*\left[J_n(u,u^*)\dfrac{\partial}{\partial u_n}\ln\dfrac{p_t(u,u^*)}{p_t({\bm U}^<_K)p_t({\bm U}^>_K)}\right.\notag\\
% &\qquad\qquad\left.+J^*_n(u,u^*)\dfrac{\partial}{\partial u^*_n}\ln\dfrac{p_t(u,u^*)}{p_t({\bm U}^<_K)p_t({\bm U}^>_K)}\right],
\label{IF_def-1}
\end{align}
Similarly, the scale-to-scale information flow from the large-scale modes ${\bm U}^<_K$ to the small-scale modes ${\bm U}^>_K$ is defined by
\begin{align}
\dot{I}^>_K[{\bm U}^<_K\colon\!{\bm U}^>_K]:=\lim_{\Delta t\rightarrow0}\dfrac{I[{\bm U}^<_K(t)\colon\!{\bm U}^>_K(t+\Delta t)]-I[{\bm U}^<_K(t)\colon\!{\bm U}^>_K(t)]}{\Delta t}.
% &=\sum^{N}_{n=n_K+1}\int dudu^*\left[J_n(u,u^*)\dfrac{\partial}{\partial u_n}\ln\dfrac{p_t(u,u^*)}{p_t({\bm U}^<_K)p_t({\bm U}^>_K)}\right.\notag\\
% &\qquad\qquad\left.+J^*_n(u,u^*)\dfrac{\partial}{\partial u^*_n}\ln\dfrac{p_t(u,u^*)}{p_t({\bm U}^<_K)p_t({\bm U}^>_K)}\right].
\label{IF_def-2}
\end{align}
Below, we shall use the simplified notation $\dot{I}^>_K=\dot{I}^>_K[{\bm U}^<_K\colon\!{\bm U}^>_K]$ and $\dot{I}^<_K=\dot{I}^<_K[{\bm U}^<_K\colon\!{\bm U}^>_K]$.
Note that the Sabra shell model (\ref{sabra shell model}) satisfies the bipartite condition because the noises acting on different shells are uncorrelated, i.e., $\langle\xi_n(t)\xi^*_{n'}(t')\rangle=0$ for $n\neq n'$.
Then, the sum of these two information flows gives the time derivative of the mutual information, as in Eq.~(\ref{dtI_Ix_Iy}):
\begin{align}
\dfrac{d}{dt}I[{\bm U}^<_K\colon\!{\bm U}^>_K]=\dot{I}^<_K+\dot{I}^>_K.
\label{dtI}
\end{align}
Therefore, in the steady state, there is only one information flow because $d_tI[{\bm U}^<_K\colon\!{\bm U}^>_K]=0$:
\begin{align}
\dot{\mathcal{I}}_K:=\dot{I}^>_K=-\dot{I}^<_K.
\label{single_IF}
\end{align}
Here, we have introduced the notation $\dot{\mathcal{I}}_K$ to denote the time-independent steady-state information flow.
Note that $\dot{\mathcal{I}}_K$ becomes zero in the equilibrium state described by Eq.~(\ref{Gibbs distribution}) because the mutual information is zero.

The sign of the scale-to-scale information flow indicates the direction of information transfer in wavenumber space.
If $\dot{\mathcal{I}}_K>0$, then it means that information about the large-scale modes is transferred to small scales (see Fig.~\ref{fig:Fourier_modes_division}).
In contrast, if $\dot{\mathcal{I}}_K<0$, it means that information is transferred from small to large scales.

In our previous paper~\cite{tanogami2024information}, by applying information thermodynamics, we proved that $\dot{\mathcal{I}}_K$ cannot be negative for $K$ within the inertial range $k_f\ll K\ll k_\nu$:
\begin{align}
\dfrac{\rho V\varepsilon}{k_{\mathrm{B}}T}\ge\dot{\mathcal{I}}_K\ge0,
\label{Information-thermodynamic bound}
\end{align}
where $V$ denotes the volume of the fluid, which does not appear in the case of the shell model because $\rho$ has units of mass.
The inequality (\ref{Information-thermodynamic bound}) states that the information of large-scale eddies is transferred to small-scale eddies along with the energy cascade.
This inequality is an exact and universal relation independent of the details of the flow under consideration.
Furthermore, this inequality is valid for various turbulence models that are thermodynamically consistent (i.e., satisfy the fluctuation-dissipation relation of the second kind) and exhibit the energy cascade, such as the fluctuating Navier--Stokes equation.

\subsection{Decomposition of the information flow into reversible and irreversible parts\label{Decomposition of the information flow into reversible and irreversible parts}}
Here, we decompose the scale-to-scale information flow into reversible and irreversible parts.
This decomposition of the information flow will be used in the proof of scale locality in the next section.

We first note that the scale-to-scale information flow can be expressed in terms of the probability current $J_n(u,u^*)$, as in Eqs.~(\ref{IF expression J_X}) and (\ref{IF expression J_Y}):
\begin{align}
\dot{I}^<_K&=\sum^{n_K}_{n=0}\int dudu^*\left[J_n(u,u^*)\dfrac{\partial}{\partial u_n}\ln\dfrac{p_t(u,u^*)}{p_t({\bm U}^<_K)p_t({\bm U}^>_K)}+\mathrm{c.c.}\right],\label{IF_expression_1}\\
\dot{I}^>_K&=\sum^{N}_{n=n_K+1}\int dudu^*\left[J_n(u,u^*)\dfrac{\partial}{\partial u_n}\ln\dfrac{p_t(u,u^*)}{p_t({\bm U}^<_K)p_t({\bm U}^>_K)}+\mathrm{c.c.}\right].\label{IF_expression_2}
% \dot{I}^<_K&=\sum^{n_K}_{n=0}\int dudu^*\left[J_n(u,u^*)\dfrac{\partial}{\partial u_n}\ln\dfrac{p_t(u,u^*)}{p_t({\bm U}^<_K)p_t({\bm U}^>_K)}+J^*_n(u,u^*)\dfrac{\partial}{\partial u^*_n}\ln\dfrac{p_t(u,u^*)}{p_t({\bm U}^<_K)p_t({\bm U}^>_K)}\right],\label{IF_expression_1}\\
% \dot{I}^>_K&=\sum^{N}_{n=n_K+1}\int dudu^*\left[J_n(u,u^*)\dfrac{\partial}{\partial u_n}\ln\dfrac{p_t(u,u^*)}{p_t({\bm U}^<_K)p_t({\bm U}^>_K)}+J^*_n(u,u^*)\dfrac{\partial}{\partial u^*_n}\ln\dfrac{p_t(u,u^*)}{p_t({\bm U}^<_K)p_t({\bm U}^>_K)}\right].\label{IF_expression_2}
\end{align}
Here, $\mathrm{c.c.}$ denotes the complex conjugate term, and $dudu^*:=\prod_nd\mathrm{Re}[u_n]d\mathrm{Im}[u_n]$, where $\mathrm{Re}[u_n]$ and $\mathrm{Im}[u_n]$ denote the real and imaginary parts of $u_n$, respectively.
These expressions can be proved using the same argument as in~\ref{Expression for information flow in terms of probability currents} (see also Appendix B of Ref.~\cite{tanogami2024information}).

We decompose the probability current $J_n(u,u^*)$ defined by Eq.~(\ref{probability current}) into two parts:
\begin{align}
J_n(u,u^*)=J^{\mathrm{rev}}_n(u,u^*)+J^{\mathrm{irr}}_n(u,u^*),
\end{align}
where $J^{\mathrm{rev}}_n(u,u^*)$ denotes the reversible current, defined by
\begin{align}
J^{\mathrm{rev}}_n(u,u^*)&:=\dfrac{1}{2}\left[J_n(u,u^*)+J_n(-u,-u^*)\right]\notag\\
&=\left(B_n(u,u^*)+f_n\right)p_t(u,u^*),
\label{reversible current}
\end{align}
and $J^{\mathrm{irr}}_n(u,u^*)$ denotes the irreversible current, defined by
\begin{align}
J^{\mathrm{irr}}_n(u,u^*)&:=\dfrac{1}{2}\left[J_n(u,u^*)-J_n(-u,-u^*)\right]\notag\\
&=-\nu k^2_nu_np_t(u,u^*)-\dfrac{2\nu k^2_nk_{\mathrm{B}}T}{\rho}\dfrac{\partial}{\partial u^*_n}p_t(u,u^*).
\label{irreversible current}
\end{align}
Note that the irreversible current consists of contributions from viscous damping and thermal fluctuations.
Using this decomposition, we can also decompose the scale-to-scale information flow into reversible and irreversible parts.
For example, $\dot{I}^<_K$ can be decomposed as follows:
\begin{align}
\dot{I}^<_K=\dot{I}^{<,\mathrm{rev}}_K+\dot{I}^{<,\mathrm{irr}}_K,
\end{align}
where
\begin{align}
\dot{I}^{<,\mathrm{rev/irr}}_K&=\sum^{n_K}_{n=0}\int dudu^*\left[J^{\mathrm{rev/irr}}_n(u,u^*)\dfrac{\partial}{\partial u_n}\ln\dfrac{p_t(u,u^*)}{p_t({\bm U}^<_K)p_t({\bm U}^>_K)}+\mathrm{c.c.}\right].
% \dot{I}^{<,\mathrm{rev/irr}}_K&=\sum^{n_K}_{n=0}\int dudu^*\left[J^{\mathrm{rev/irr}}_n(u,u^*)\dfrac{\partial}{\partial u_n}\ln\dfrac{p_t(u,u^*)}{p_t({\bm U}^<_K)p_t({\bm U}^>_K)}\right.\notag\\
% &\qquad\qquad\left.+J^{\mathrm{rev/irr}*}_n(u,u^*)\dfrac{\partial}{\partial u^*_n}\ln\dfrac{p_t(u,u^*)}{p_t({\bm U}^<_K)p_t({\bm U}^>_K)}\right].
% &=\sum^{n_K}_{n=0}\int dudu^*\left[\left(B_n(u,u^*)+f_n\right)p_t(u,u^*)\dfrac{\partial}{\partial u_n}\ln\dfrac{p_t(u,u^*)}{p_t({\bm U}^<_K)p_t({\bm U}^>_K)}+\mathrm{c.c.}\right],
\end{align}
We note that, in the equilibrium state described by Eq.~(\ref{Gibbs distribution}), both $\dot{I}^{<,\mathrm{rev}}_K$ and $\dot{I}^{<,\mathrm{irr}}_K$ are zero, although the reversible current $J^{\mathrm{rev}}_n(u,u^*)$ is generally nonzero.

The important point here is that, in the fully developed turbulent regime, the irreversible part $\dot{I}^{<,\mathrm{irr}}_K$ can be ignored in the inertial range $k_f\ll K\ll k_\nu$ because the viscous and thermal noise effects are negligible:
\begin{align}
\dot{I}^<_K=\dot{I}^{<,\mathrm{rev}}_K \qquad \text{for}\quad k_f\ll K\ll k_\nu.
\label{IF=IF_rev}
\end{align}
For a more detailed derivation of Eq.~(\ref{IF=IF_rev}) using a dimensionless form, see~\ref{Detailed discussion on irreversible information flow}.
% Similarly to $\dot{I}^<_K$, the information flow $\dot{I}^>_K$ can also be decomposed as $\dot{I}^>_K=\dot{I}^{>,\mathrm{rev}}_K+\dot{I}^{>,\mathrm{irr}}_K$.
% In contrast to $\dot{I}^{<,\mathrm{irr}}_K$, the irreversible part $\dot{I}^{>,\mathrm{irr}}_K$ cannot be ignored because it contains the high-wave-number shell variables.
% \textbf{However, since $\dot{\mathcal{I}}_K=\dot{I}^>_K=-\dot{I}^<_K$ in the steady state, the scale-to-scale information flow is governed by the nonlinear interaction rather than the viscous stress and thermal fluctuations.}
Equation (\ref{IF=IF_rev}) can be further simplified by noting that the external force $f_n$ in the reversible probability current (\ref{reversible current}) does not contribute to the reversible part $\dot{I}^{<,\mathrm{rev}}_K$.
In fact,
\begin{align}
&\quad\sum^{n_K}_{n=0}\int dudu^*\left[f_np_t(u,u^*)\dfrac{\partial}{\partial u_n}\ln\dfrac{p_t(u,u^*)}{p_t({\bm U}^<_K)p_t({\bm U}^>_K)}+\mathrm{c.c.}\right]\notag\\
&=\sum^{n_K}_{n=0}\int dudu^*\left[f_n\left(\dfrac{\partial}{\partial u_n}p_t(u,u^*)-p_t({\bm U}^>_K|{\bm U}^<_K)\dfrac{\partial}{\partial u_n}p_t({\bm U}^<_K)\right)+\mathrm{c.c.}\right]\notag\\
&=\sum^{n_K}_{n=0}\left[f_n\left(\int du_n\dfrac{\partial}{\partial u_n}p_t(u_n)-\int du_n\dfrac{\partial}{\partial u_n}p_t(u_n)\right)+\mathrm{c.c.}\right]\notag\\
&=0,
\end{align}
where $p_t({\bm U}^>_K|{\bm U}^<_K)=p_t(u,u^*)/p_t({\bm U}^<_K)$ denotes the conditional probability density, and $p_t(u_n)$ denotes the marginal distribution of $u_n$.
Therefore, for $K$ within the inertial range, the scale-to-scale information flow $\dot{I}^<_K$ can be expressed as
\begin{align}
\dot{I}^<_K&=\sum^{n_K}_{n=0}\int dudu^*\left[B_n(u,u^*)p_t(u,u^*)\dfrac{\partial}{\partial u_n}\ln\dfrac{p_t(u,u^*)}{p_t({\bm U}^<_K)p_t({\bm U}^>_K)}+\mathrm{c.c.}\right].
\label{IF in terms of reversible current}
\end{align}
This expression implies that the scale-to-scale information flow in turbulence is governed by the nonlinear interactions rather than thermal fluctuations.
In other words, the nature of the scale-to-scale information flow in the inertial range may be essentially the same for the deterministic and stochastic cases, as implied by the numerical simulation in Ref.~\cite{tanogami2024information}.

\section{Scale locality of the information flow\label{Scale locality of the information flow}} 
In this section, we demonstrate that the scale-to-scale information flow satisfies the scale locality in the inertial range.
We first introduce the scale-local and nonlocal modes in Sec.~\ref{subsec: Scale-local and nonlocal modes}.
Then, we show that the scale-to-scale information flow can be decomposed into scale-local and scale-nonlocal parts in Sec.~\ref{subsec: Decomposition of the information flow into scale-local and scale-nonlocal parts}.
Based on this decomposition, we prove that the scale-nonlocal part can be ignored compared to the scale-local part in Sec.~\ref{subsec: Scale locality of the information flow}.
% Because the proof is rather long and technical, we provide an outline of the proof in this section.
% See~\ref{Proof of the scale locality} for a detailed derivation.

\subsection{Scale-local and nonlocal modes\label{subsec: Scale-local and nonlocal modes}}
\begin{figure*}[t]
% \center
\centering
\includegraphics[width=16cm]{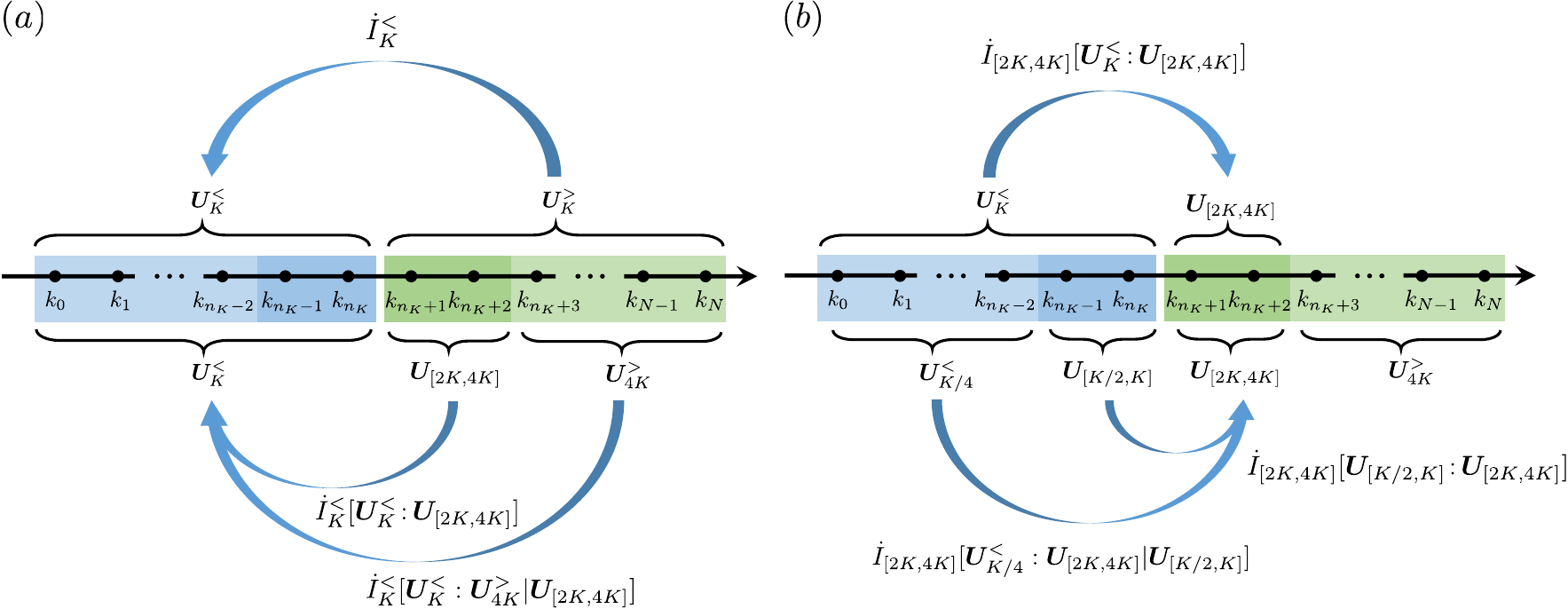}
\caption{Schematic of band-pass, low-pass, and high-pass filtered shell variables and the decomposition of the scale-to-scale information flow into scale-local and scale-nonlocal parts. 
(a) The decomposition of $\dot{I}^<_K$ into the sum of $\dot{I}^<_K[{\bm U}^<_K\colon\!{\bm U}_{[2K,4K]}]$ and $\dot{I}^<_K[{\bm U}^<_K\colon\!{\bm U}^>_{4K}|{\bm U}_{[2K,4K]}]$.
(b) The decomposition of $\dot{I}_{[2K,4K]}[{\bm U}^<_K\colon\!{\bm U}_{[2K,4K]}]$ into the sum of $\dot{I}_{[2K,4K]}[{\bm U}_{[K/2,K]}\colon\!{\bm U}_{[2K,4K]}]$ and $\dot{I}_{[2K,4K]}[{\bm U}^<_{K/4}\colon\!{\bm U}_{[2K,4K]}|{\bm U}_{[K/2,K]}]$.
In both panels, the arrow indicates the direction of the corresponding information flow when it is positive.}
\label{fig:schematic_proof_decomposition}
\end{figure*}

First, we introduce important quantities that characterize the scale locality.
We define the scale-local and scale-nonlocal shell variables at scale $K$, which allows us to further decompose the large-scale and small-scale modes ${\bm U}^<_K$ and ${\bm U}^>_K$.
In the following, we assume that $K$ is within the inertial range $k_f\ll K\ll k_\nu$.
The scale-local modes are defined as band-pass filtered shell variables as follows:
\begin{align}
{\bm U}_{[K/2,K]}&:=\{u_{n_K-1},u^*_{n_K-1},u_{n_K},u^*_{n_K}\},\\
{\bm U}_{[2K,4K]}&:=\{u_{n_K+1},u^*_{n_K+1},u_{n_K+2},u^*_{n_K+2}\}.
\end{align}
Here, note that $K/2=k_{n_K-1}$, $K=k_{n_K}$, $2K=k_{n_K+1}$, and $4K=k_{n_K+2}$.
Similarly, the scale-nonlocal modes are defined as low-pass and high-pass filtered shell variables:
\begin{align}
{\bm U}^<_{K/4}&:=\{u_n,u^*_n\mid0\le n\le n_K-2\},\\
{\bm U}^>_{4K}&:=\{u_n,u^*_n\mid n_K+2<n\le N\}.
\end{align}
See Fig.~\ref{fig:schematic_proof_decomposition} for a schematic of the scale-local and nonlocal modes.
From the definition of these scale-local and scale-nonlocal modes, the large-scale and small-scale modes ${\bm U}^<_K$ and ${\bm U}^>_K$ can be decomposed as follows:
\begin{align}
{\bm U}^<_K={\bm U}^<_{K/4}\cup{\bm U}_{[K/2,K]},\quad {\bm U}^>_K={\bm U}_{[2K,4K]}\cup{\bm U}^>_{4K}.
\label{decomposition of large and small scale modes}
\end{align}
By noting that the direct nonlinear interaction $B_n(u,u^*)$ is limited to the nearest- and next-nearest-neighbor shells, this decomposition can also be interpreted as a decomposition into direct and indirect interaction parts.
Indeed, the scale-local modes ${\bm U}_{[K/2,K]}$ is a subset of ${\bm U}^<_K$ that interacts directly with ${\bm U}^>_K$ while there is no direct coupling between the scale-nonlocal modes ${\bm U}^<_{K/4}$ and ${\bm U}^>_K$.
Similarly, ${\bm U}_{[2K,4K]}$ is a subset of ${\bm U}^>_K$ that interacts directly with ${\bm U}^<_K$ while there is no direct coupling between ${\bm U}^>_{4K}$ and ${\bm U}^<_K$.
Note that the results described below remain unchanged if we instead define the scale-local modes with $m$ shells ($2\le m\ll N$).
That is,
\begin{align}
{\bm U}^<_K&={\bm U}^<_{k_{n_K-m}}\cup{\bm U}_{[k_{n_K-m+1},k_{n_K}]},\label{m-decomposition of large and small scale modes_1}\\
{\bm U}^>_K&={\bm U}_{[k_{n_K+1},k_{n_K+m}]}\cup{\bm U}^>_{k_{n_K+m}}.
\label{m-decomposition of large and small scale modes_2}
\end{align}
Here, the integer $m$ is restricted to be greater than or equal to $2$ so that we can decompose the shell variables into direct and indirect interaction parts.
The proof of the scale locality presented in this paper does not directly apply to the decomposition (\ref{m-decomposition of large and small scale modes_1}) and (\ref{m-decomposition of large and small scale modes_2}) with $m=1$.

\subsection{Decomposition of the information flow into scale-local and scale-nonlocal parts\label{subsec: Decomposition of the information flow into scale-local and scale-nonlocal parts}}
By using the scale-local and scale-nonlocal modes and the chain rule for mutual information~\cite{cover1999elements,horowitz2015multipartite}, we can decompose the steady-state scale-to-scale information flow $\dot{\mathcal{I}}_K$ into scale-local and scale-nonlocal parts.
First, by using the decomposition of the shell variables (\ref{decomposition of large and small scale modes}), we note that 
\begin{align}
\ln\dfrac{p_t({\bm U}^<_K,{\bm U}^>_K)}{p_t({\bm U}^<_K)p_t({\bm U}^>_K)}=\ln\dfrac{p_t({\bm U}^<_K,{\bm U}_{[2K,4K]})}{p_t({\bm U}^<_K)p_t({\bm U}_{[2K,4K]})}+\ln\dfrac{p_t({\bm U}^<_K,{\bm U}^>_{4K}|{\bm U}_{[2K,4K]})}{p_t({\bm U}^<_K|{\bm U}_{[2K,4K]})p_t({\bm U}^>_{4K}|{\bm U}_{[2K,4K]})}.
\end{align}
From this chain rule and Eq.~(\ref{IF_expression_1}), we obtain (see Fig.~\ref{fig:schematic_proof_decomposition}(a))
\begin{align}
\dot{\mathcal{I}}_K&=-\dot{I}^<_K\notag\\
&=-\dot{I}^<_K[{\bm U}^<_K\colon\!{\bm U}_{[2K,4K]}]-\dot{I}^<_K[{\bm U}^<_K\colon\!{\bm U}^>_{4K}|{\bm U}_{[2K,4K]}],
\label{ultraviolet decomposition}
\end{align}
where $\dot{I}^<_K[{\bm U}^<_K\colon\!{\bm U}_{[2K,4K]}]$ denotes the information flow from ${\bm U}_{[2K,4K]}$ to ${\bm U}^<_K$:
\begin{align}
\dot{I}^<_K[{\bm U}^<_K\colon\!{\bm U}_{[2K,4K]}]&:=\lim_{\Delta t\rightarrow0}\dfrac{1}{\Delta t}\biggl(I[{\bm U}^<_K(t+\Delta t)\colon\!{\bm U}_{[2K,4K]}(t)]-I[{\bm U}^<_K(t)\colon\!{\bm U}_{[2K,4K]}(t)]\biggr)\notag\\
&=\sum^{n_K}_{n=0}\int dudu^*\left[J_n(u,u^*)\dfrac{\partial}{\partial u_n}\ln\dfrac{p_t({\bm U}^<_K,{\bm U}_{[2K,4K]})}{p_t({\bm U}^<_K)p_t({\bm U}_{[2K,4K]})}+\mathrm{c.c.}\right].
\end{align}
In the second line, we expressed the information flow in terms of the probability current, as in Eqs.~(\ref{IF_expression_1}) and (\ref{IF_expression_2}).
These expressions can be proved using the same argument as in~\ref{Expression for information flow in terms of probability currents} (the same applies to the last lines of Eqs.~(\ref{scale-nonlocal information flow-2}), (\ref{UV local information flow}), (\ref{scale-local information flow}), and (\ref{scale-nonlocal information flow-1}) below).
The second term on the right-hand side of Eq.~(\ref{ultraviolet decomposition}) denotes the information flow from ${\bm U}^>_{4K}$ to ${\bm U}^<_K$ under the condition of ${\bm U}_{[2K,4K]}$ defined by
\begin{align}
&\qquad\dot{I}^<_K[{\bm U}^<_K\colon\!{\bm U}^>_{4K}|{\bm U}_{[2K,4K]}]\notag\\
&:=\lim_{\Delta t\rightarrow0}\dfrac{1}{\Delta t}\biggl(I[{\bm U}^<_K(t+\Delta t)\colon\!{\bm U}^>_{4K}(t)|{\bm U}_{[2K,4K]}(t)]-I[{\bm U}^<_K(t)\colon\!{\bm U}^>_{4K}(t)|{\bm U}_{[2K,4K]}(t)]\biggr)\notag\\
&=\sum^{n_K}_{n=0}\int dudu^*\left[J_n(u,u^*)\dfrac{\partial}{\partial u_n}\ln\dfrac{p_t({\bm U}^<_K,{\bm U}^>_{4K}|{\bm U}_{[2K,4K]})}{p_t({\bm U}^<_K|{\bm U}_{[2K,4K]})p_t({\bm U}^>_{4K}|{\bm U}_{[2K,4K]})}+\mathrm{c.c.}\right].
\label{scale-nonlocal information flow-2}
\end{align}

The information flow $\dot{I}^<_K[{\bm U}^<_K\colon\!{\bm U}_{[2K,4K]}]$ can be further decomposed as follows.
First, we introduce the information flow from ${\bm U}^<_K$ to ${\bm U}_{[2K,4K]}$ defined by (see also Fig.~\ref{fig:schematic_proof_decomposition}(b))
\begin{align}
\dot{I}_{[2K,4K]}[{\bm U}^<_K\colon\!{\bm U}_{[2K,4K]}]&:=\lim_{\Delta t\rightarrow0}\dfrac{1}{\Delta t}\biggl(I[{\bm U}^<_K(t)\colon\!{\bm U}_{[2K,4K]}(t+\Delta t)]-I[{\bm U}^<_K(t)\colon\!{\bm U}_{[2K,4K]}(t)]\biggr)\notag\\
&=\sum^{n_K+2}_{n=n_K+1}\int dudu^*\left[J_n(u,u^*)\dfrac{\partial}{\partial u_n}\ln\dfrac{p_t({\bm U}^<_K,{\bm U}_{[2K,4K]})}{p_t({\bm U}^<_K)p_t({\bm U}_{[2K,4K]})}+\mathrm{c.c.}\right].
\label{UV local information flow}
\end{align}
This information flow is related to $\dot{I}^<_K[{\bm U}^<_K\colon\!{\bm U}_{[2K,4K]}]$ via the following relation, as in Eq.~(\ref{dtI_Ix_Iy}):
\begin{align}
\dfrac{d}{dt}I[{\bm U}^<_K\colon\!{\bm U}_{[2K,4K]}]=\dot{I}^<_K[{\bm U}^<_K\colon\!{\bm U}_{[2K,4K]}]+\dot{I}_{[2K,4K]}[{\bm U}^<_K\colon\!{\bm U}_{[2K,4K]}].
\label{stationarity_ultraviolet local IF}
\end{align}
Hence, we find that
\begin{align}
\dot{I}^<_K[{\bm U}^<_K\colon\!{\bm U}_{[2K,4K]}]=-\dot{I}_{[2K,4K]}[{\bm U}^<_K\colon\!{\bm U}_{[2K,4K]}]
\label{UV local information flow_steady state}
\end{align}
in the steady state.
From the chain rule
\begin{align}
\ln\dfrac{p_t({\bm U}^<_K,{\bm U}_{[2K,4K]})}{p_t({\bm U}^<_K)p_t({\bm U}_{[2K,4K]})}&=\ln\dfrac{p_t({\bm U}_{[K/2,K]},{\bm U}_{[2K,4K]})}{p_t({\bm U}_{[K/2,K]})p_t({\bm U}_{[2K,4K]})}\notag\\
&\qquad+\ln\dfrac{p_t({\bm U}^<_{K/4},{\bm U}_{[2K,4K]}|{\bm U}_{[K/2,K]})}{p_t({\bm U}^<_{K/4}|{\bm U}_{[K/2,K]})p_t({\bm U}_{[2K,4K]}|{\bm U}_{[K/2,K]})},
\end{align}
we can decompose $\dot{I}_{[2K,4K]}[{\bm U}^<_K\colon\!{\bm U}_{[2K,4K]}]$ as follows (see Fig.~\ref{fig:schematic_proof_decomposition}(b)):
\begin{align}
\dot{I}_{[2K,4K]}[{\bm U}^<_K\colon\!{\bm U}_{[2K,4K]}]=\dot{I}_{[2K,4K]}[{\bm U}_{[K/2,K]}\colon\!{\bm U}_{[2K,4K]}]+\dot{I}_{[2K,4K]}[{\bm U}^<_{K/4}\colon\!{\bm U}_{[2K,4K]}|{\bm U}_{[K/2,K]}],
\label{Infrared decomposition}
\end{align}
where $\dot{I}_{[2K,4K]}[{\bm U}_{[K/2,K]}\colon\!{\bm U}_{[2K,4K]}]$ denotes the information flow from ${\bm U}_{[K/2,K]}$ to ${\bm U}_{[2K,4K]}$,
\begin{align}
% \dot{\mathcal{I}}^{\mathrm{local}}_K&\equiv
&\qquad\dot{I}_{[2K,4K]}[{\bm U}_{[K/2,K]}\colon\!{\bm U}_{[2K,4K]}]\notag\\
&:=\lim_{\Delta t\rightarrow0}\dfrac{1}{\Delta t}\biggl(I[{\bm U}_{[K/2,K]}(t)\colon\!{\bm U}_{[2K,4K]}(t+\Delta t)]-I[{\bm U}_{[K/2,K]}(t)\colon\!{\bm U}_{[2K,4K]}(t)]\biggr)\notag\\
&=\sum^{n_K+2}_{n=n_K+1}\int dudu^*\left[J_n(u,u^*)\dfrac{\partial}{\partial u_n}\ln\dfrac{p_t({\bm U}_{[K/2,K]},{\bm U}_{[2K,4K]})}{p_t({\bm U}_{[K/2,K]})p_t({\bm U}_{[2K,4K]})}+\mathrm{c.c.}\right],
\label{scale-local information flow}
\end{align}
and $\dot{I}_{[2K,4K]}[{\bm U}^<_{K/4}\colon\!{\bm U}_{[2K,4K]}|{\bm U}_{[K/2,K]}]$ denotes the information flow from ${\bm U}^<_{K/4}$ to ${\bm U}_{[2K,4K]}$ under the condition of ${\bm U}_{[K/2,K]}$,
\begin{align}
&\quad\dot{I}_{[2K,4K]}[{\bm U}^<_{K/4}\colon\!{\bm U}_{[2K,4K]}|{\bm U}_{[K/2,K]}]\notag\\
&:=\lim_{\Delta t\rightarrow0}\dfrac{1}{\Delta t}\biggl(I[{\bm U}^<_{K/4}(t)\colon\!{\bm U}_{[2K,4K]}(t+\Delta t)|{\bm U}_{[K/2,K]}(t)]-I[{\bm U}^<_{K/4}(t)\colon\!{\bm U}_{[2K,4K]}(t)|{\bm U}_{[K/2,K]}(t)]\biggr)\notag\\
&=\sum^{n_K+2}_{n=n_K+1}\int dudu^*\left[J_n(u,u^*)\dfrac{\partial}{\partial u_n}\ln\dfrac{p_t({\bm U}^<_{K/4},{\bm U}_{[2K,4K]}|{\bm U}_{[K/2,K]})}{p_t({\bm U}^<_{K/4}|{\bm U}_{[K/2,K]})p_t({\bm U}_{[2K,4K]}|{\bm U}_{[K/2,K]})}+\mathrm{c.c.}\right].
\label{scale-nonlocal information flow-1}
\end{align}
% where $p_t(\cdot|{\bm U}_{[K/2,K]})$ denotes the probability density under the condition of ${\bm U}_{[K/2,K]}$.
% By combining Eqs.~(\ref{ultraviolet decomposition}), (\ref{stationarity_ultraviolet local IF}), and (\ref{Infrared decomposition}), we arrive at Eq.~(\ref{decomposition of information flow}).

\begin{figure*}[t]
\center
\centering
\includegraphics[width=16cm]{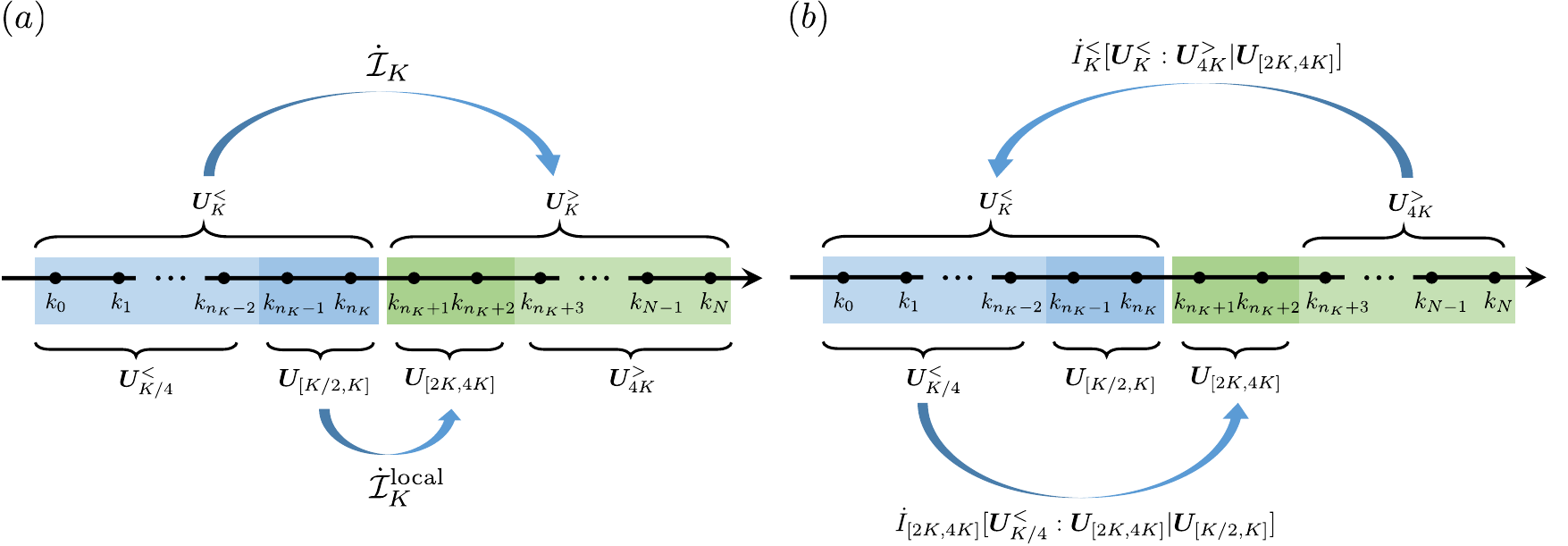}
\caption{Schematic of the decomposition of the scale-to-scale information flow into scale-local and scale-nonlocal parts. 
(a) The original scale-to-scale information flow $\dot{\mathcal{I}}_K$ and the scale-local information flow $\dot{\mathcal{I}}^{\mathrm{local}}_K$.
(b) The scale-nonlocal contribution to the information flow: $\dot{I}_{[2K,4K]}[{\bm U}^<_{K/4}\colon\!{\bm U}_{[2K,4K]}|{\bm U}_{[K/2,K]}]$ and $\dot{I}^<_K[{\bm U}^<_K\colon\!{\bm U}^>_{4K}|{\bm U}_{[2K,4K]}]$.
In both panels, the arrow indicates the direction of the corresponding information flow when it is positive.}
\label{fig:shell_filtering}
\end{figure*}

By combining Eqs.~(\ref{ultraviolet decomposition}), (\ref{UV local information flow_steady state}), and (\ref{Infrared decomposition}), we finally obtain
\begin{align}
\dot{\mathcal{I}}_K&=\dot{I}_{[2K,4K]}[{\bm U}_{[K/2,K]}\colon\!{\bm U}_{[2K,4K]}]\notag\\
&\qquad+\dot{I}_{[2K,4K]}[{\bm U}^<_{K/4}\colon\!{\bm U}_{[2K,4K]}|{\bm U}_{[K/2,K]}]-\dot{I}^<_K[{\bm U}^<_K\colon\!{\bm U}^>_{4K}|{\bm U}_{[2K,4K]}]\notag\\
&=:\dot{\mathcal{I}}^{\mathrm{local}}_K+\dot{\mathcal{I}}^{\mathrm{nonlocal}}_K,
\label{scale-local and nonlocal decomposition}
\end{align}
where 
\begin{align}
\dot{\mathcal{I}}^{\mathrm{local}}_K&:=\dot{I}_{[2K,4K]}[{\bm U}_{[K/2,K]}\colon\!{\bm U}_{[2K,4K]}]\,\Bigl(=-\dot{I}_{[K/2,K]}[{\bm U}_{[K/2,K]}\colon\!{\bm U}_{[2K,4K]}]\Bigr)
\label{def: scale-local information flow}
\end{align}
denotes the steady-state scale-local information flow, and
\begin{align}
\dot{\mathcal{I}}^{\mathrm{nonlocal}}_K:=\dot{I}_{[2K,4K]}[{\bm U}^<_{K/4}\colon\!{\bm U}_{[2K,4K]}|{\bm U}_{[K/2,K]}]-\dot{I}^<_K[{\bm U}^<_K\colon\!{\bm U}^>_{4K}|{\bm U}_{[2K,4K]}].
\label{def: scale-nonlocal information flow}
\end{align}
denotes the steady-state scale-nonlocal information flow.
See Fig.~\ref{fig:shell_filtering} for a schematic of the scale-local and nonlocal information flow.
The decomposition (\ref{scale-local and nonlocal decomposition}) is the key result in proving the scale locality of the information flow.
We remark that the decomposition (\ref{scale-local and nonlocal decomposition}) can be interpreted as a decomposition of $\dot{\mathcal{I}}_K$ into an information flow between directly interacting subsystems and that between subsystems that do not interact directly.
Such a decomposition was first employed in Ref.~\cite{horowitz2015multipartite} in the context of multiple Maxwell demons.

\subsection{Scale locality of the information flow\label{subsec: Scale locality of the information flow}}
In the inertial range, we can show that the scale-nonlocal contribution $\dot{\mathcal{I}}^{\mathrm{nonlocal}}_K$ can be ignored compared to $\dot{\mathcal{I}}^{\mathrm{local}}_K$.
That is, under the Kolmogorov hypothesis for the Kolmogorov multiplier~\cite{kolmogorov1962refinement}, the scale-to-scale information flow $\dot{\mathcal{I}}_K$ is approximately equal to the scale-local information flow $\dot{\mathcal{I}}^{\mathrm{local}}_K$:
\begin{align}
\dot{\mathcal{I}}_K\simeq\dot{\mathcal{I}}^{\mathrm{local}}_K\qquad \text{for}\quad k_f\ll K\ll k_\nu,
\label{scale locality}
\end{align}
which is the central result of this paper.
This relation states that the information transfer from large to small scales occurs mainly through scale-local interactions, which is consistent with the scale locality of the energy cascade.
% Here, we have used the approximation that the conditional probability density $p_t({\bm U}^>_{4K}|{\bm U}^<_{4K})$ has a short-range dependence on the large-scale modes ${\bm U}^<_{4K}$, i.e., $p_t({\bm U}^>_{4K}|{\bm U}^<_{4K})\simeq p_t({\bm U}^>_{4K}|{\bm U}_{[K/2,4K]})$, where ${\bm U}_{[K/2,4K]}={\bm U}_{[K/2,K]}\cup{\bm U}_{[2K,4K]}$.
% This approximation is based on Kolmogorov's third hypothesis~\cite{kolmogorov1962refinement,chen2003kolmogorov,eyink2003gibbsian}, and its validity is verified numerically in Ref.~\cite{biferale2017optimal}.
In other words, turbulent fluctuations at large scales do not directly influence those at small scales without scale-local interactions.

% Remarks:
% Here, we provide a brief overview of the proof of Eq.~(\ref{scale locality}).
% The proof is based on the expression of the information flow in terms of the reversible current [Eq.~(\ref{IF in terms of reversible current})] and the scale locality of the nonlinear interaction $B_n(u,u^*)$.
% Specifically, the proof consists of two steps: (i) $\dot{I}^<_K[{\bm U}^<_K\colon\!{\bm U}^>_{4K}|{\bm U}_{[2K,4K]}]=0$ and (ii) $\dot{I}_{[2K,4K]}[{\bm U}^<_{K/4}\colon\!{\bm U}_{[2K,4K]}|{\bm U}_{[K/2,K]}]=0$.
% Note that the part (i) can be interpreted as a condition for ultraviolet (UV) locality because it means that there is no direct influence from high-wavenumber modes ${\bm U}^>_{4K}$ to ${\bm U}^<_K$, in particular to ${\bm U}_{[K/2,K]}$.
% Similarly, the part (ii) can be interpreted as a condition for infrared (IR) locality because it means that there is no direct influence from low-wavenumber modes ${\bm U}^<_{K/4}$ to ${\bm U}_{[2K,4K]}$.
% The UV locality can be proved simply by noting that there is no direct interaction between ${\bm U}^<_K$ and ${\bm U}^>_{4K}$.
% The proof of the IR locality is slightly complicated because an effective nonlinear interaction appears, which is not strictly scale-local.
% Even in this case, if we assume the Kolmogorov hypothesis, the effective nonlinear interaction can be approximated as a scale-local interaction, and thus, we can prove the scale locality of the information flow.
% For a detailed derivation, see~\ref{Proof of the scale locality}.

Below, we prove the scale locality~(\ref{scale locality}).
The proof of the scale locality consists of two steps: (i) $\dot{I}^<_K[{\bm U}^<_K\colon\!{\bm U}^>_{4K}|{\bm U}_{[2K,4K]}]=0$ and (ii) $\dot{I}_{[2K,4K]}[{\bm U}^<_{K/4}\colon\!{\bm U}_{[2K,4K]}|{\bm U}_{[K/2,K]}]\simeq0$.
Note that the part (i) can be interpreted as a condition for \textit{ultraviolet locality} because it means that there is no direct influence from high-wavenumber modes ${\bm U}^>_{4K}$ to ${\bm U}^<_K$, in particular to ${\bm U}_{[K/2,K]}$.
Similarly, the part (ii) can be interpreted as a condition for \textit{infrared locality} because it means that there is no direct influence from low-wavenumber modes ${\bm U}^<_{K/4}$ to ${\bm U}_{[2K,4K]}$.
We remark that whereas the ultraviolet locality holds exactly, the infrared locality is only demonstrated approximately under the Kolmogorov hypothesis.

\subsubsection{Ultraviolet locality \label{Ultraviolet locality}}
Here, we prove the ultraviolet locality.
Our starting point is the expression of $\dot{I}^<_K[{\bm U}^<_K\colon\!{\bm U}^>_{4K}|{\bm U}_{[2K,4K]}]$ in terms of the probability current [Eq.~(\ref{scale-nonlocal information flow-2})]:
\begin{align}
&\dot{I}^<_K[{\bm U}^<_K\colon\!{\bm U}^>_{4K}|{\bm U}_{[2K,4K]}]\notag\\
&=\sum^{n_K}_{n=0}\int dudu^*\left[J_n(u,u^*)\dfrac{\partial}{\partial u_n}\ln\dfrac{p_t({\bm U}^<_K,{\bm U}^>_{4K}|{\bm U}_{[2K,4K]})}{p_t({\bm U}^<_K|{\bm U}_{[2K,4K]})p_t({\bm U}^>_{4K}|{\bm U}_{[2K,4K]})}+\mathrm{c.c.}\right].
% \label{scale-nonlocal information flow-2}
\end{align}
For $K$ within the inertial range, because the viscous and thermal noise terms in $J_n(u,u^*)$ can be ignored and the contribution from $f_n$ vanishes (see Sec.~\ref{Decomposition of the information flow into reversible and irreversible parts}), this conditional information flow can be further expressed as follows:
\begin{align}
&\dot{I}^<_K[{\bm U}^<_K\colon\!{\bm U}^>_{4K}|{\bm U}_{[2K,4K]}]\notag\\
&=\sum^{n_K}_{n=0}\int dudu^*\left[B_n(u,u^*)p_t(u,u^*)\dfrac{\partial}{\partial u_n}\ln\dfrac{p_t({\bm U}^<_K,{\bm U}^>_{4K}|{\bm U}_{[2K,4K]})}{p_t({\bm U}^<_K|{\bm U}_{[2K,4K]})p_t({\bm U}^>_{4K}|{\bm U}_{[2K,4K]})}+\mathrm{c.c.}\right].
\label{scale-nonlocal information flow_reversible current expression}
\end{align}
From the relation ${\bm U}^<_{4K}={\bm U}^<_{K}\cup{\bm U}_{[2K,4K]}$, it follows that
\begin{align}
\dfrac{p_t({\bm U}^<_K,{\bm U}^>_{4K}|{\bm U}_{[2K,4K]})}{p_t({\bm U}^<_K|{\bm U}_{[2K,4K]})}=p_t({\bm U}^>_{4K}|{\bm U}^<_{4K}).
\label{conditional probability relation}
\end{align}
By substituting Eq.~(\ref{conditional probability relation}) into Eq.~(\ref{scale-nonlocal information flow_reversible current expression}), we obtain
\begin{align}
\dot{I}^<_K[{\bm U}^<_K\colon\!{\bm U}^>_{4K}|{\bm U}_{[2K,4K]}]&=\sum^{n_K}_{n=0}\int dudu^*\left[B_n(u,u^*)p_t(u,u^*)\dfrac{\partial}{\partial u_n}\ln\dfrac{p_t({\bm U}^>_{4K}|{\bm U}^<_{4K})}{p_t({\bm U}^>_{4K}|{\bm U}_{[2K,4K]})}+\mathrm{c.c.}\right]\notag\\
&=\sum^{n_K}_{n=0}\int dudu^*\left[B_n(u,u^*)p_t({\bm U}^<_{4K})\dfrac{\partial}{\partial u_n}p_t({\bm U}^>_{4K}|{\bm U}^<_{4K})+\mathrm{c.c.}\right],
\label{UV locality}
\end{align}
where we used the fact that $\frac{\partial}{\partial u_n}p_t({\bm U}^>_{4K}|{\bm U}_{[2K,4K]})=0$ for $n=0,\ldots,n_K$.
Because $B_n(u,u^*)$ does not depend on ${\bm U}^>_{4K}$ for $n\le n_K$, i.e., $B_n(u,u^*)=B_n({\bm U}^<_{4K})$, we can show that the right-hand side of Eq.~(\ref{UV locality}) is zero:
\begin{align}
\dot{I}^<_K[{\bm U}^<_K\colon\!{\bm U}^>_{4K}|{\bm U}_{[2K,4K]}]&=\sum^{n_K}_{n=0}\int d{\bm U}^<_{4K}\left[B_n({\bm U}^<_{4K})p_t({\bm U}^<_{4K})\dfrac{\partial}{\partial u_n}\int d{\bm U}^>_{4K}p_t({\bm U}^>_{4K}|{\bm U}^<_{4K})+\mathrm{c.c.}\right]\notag\\
&=0,
% \label{UV locality}
\end{align}
where we used $\frac{\partial}{\partial u_n}\int d{\bm U}^>_{4K}p_t({\bm U}^>_{4K}|{\bm U}^<_{4K})=\frac{\partial}{\partial u_n}1=0$ in the last line.
Thus, the ultraviolet locality holds exactly.

\subsubsection{Infrared locality\label{Infrared locality}}
Next, we prove the infrared locality.
We first note that $\dot{I}_{[2K,4K]}[{\bm U}^<_{K/4}\colon\!{\bm U}_{[2K,4K]}|{\bm U}_{[K/2,K]}]$ can be expressed in terms of the probability current as follows [Eq.~(\ref{scale-nonlocal information flow-1})]:
\begin{align}
&\quad\dot{I}_{[2K,4K]}[{\bm U}^<_{K/4}\colon\!{\bm U}_{[2K,4K]}|{\bm U}_{[K/2,K]}]\notag\\
&=\sum^{n_K+2}_{n=n_K+1}\int dudu^*\left[J_n(u,u^*)\dfrac{\partial}{\partial u_n}\ln\dfrac{p_t({\bm U}^<_{K/4},{\bm U}_{[2K,4K]}|{\bm U}_{[K/2,K]})}{p_t({\bm U}^<_{K/4}|{\bm U}_{[K/2,K]})p_t({\bm U}_{[2K,4K]}|{\bm U}_{[K/2,K]})}+\mathrm{c.c.}\right].
\end{align}
For $K$ within the inertial range, because the viscous and thermal noise terms in $J_n(u,u^*)$ can be ignored and the contribution from $f_n$ vanishes (see Sec.~\ref{Decomposition of the information flow into reversible and irreversible parts}), we can show that
\begin{align}
&\quad\dot{I}_{[2K,4K]}[{\bm U}^<_{K/4}\colon\!{\bm U}_{[2K,4K]}|{\bm U}_{[K/2,K]}]\notag\\
&=\sum^{n_K+2}_{n=n_K+1}\int dudu^*\left[B_n(u,u^*)p_t(u,u^*)\dfrac{\partial}{\partial u_n}\ln\dfrac{p_t({\bm U}^<_{K/4},{\bm U}_{[2K,4K]}|{\bm U}_{[K/2,K]})}{p_t({\bm U}^<_{K/4}|{\bm U}_{[K/2,K]})p_t({\bm U}_{[2K,4K]}|{\bm U}_{[K/2,K]})}+\mathrm{c.c.}\right].
% &=\sum^{n_K+2}_{n=n_K+1}\int dudu^*\left[B_n(u,u^*)p_t(u,u^*)\dfrac{\partial}{\partial u_n}\ln\dfrac{p_t({\bm U}^<_{K/4}|{\bm U}_{[K/2,4K]})}{p_t({\bm U}^<_{K/4}|{\bm U}_{[K/2,K]})}+\mathrm{c.c.}\right]\notag\\
% &=\sum^{n_K+2}_{n=n_K+1}\int dudu^*\left[B_n(u,u^*)p_t({\bm U}^>_{4K}|{\bm U}^<_{4K})p_t({\bm U}_{[K/2,4K]})\dfrac{\partial}{\partial u_n}p_t({\bm U}^<_{K/4}|{\bm U}_{[K/2,4K]})+\mathrm{c.c.}\right].
% &=\sum^{n_K+2}_{n=n_K+1}\int d{\bm U}^<_{K/4}d{\bm U}_{[K/2,4K]}\left[\overline{B}_n({\bm U}^<_{4K})p_t({\bm U}_{[K/2,4K]})\dfrac{\partial}{\partial u_n}p_t({\bm U}^<_{K/4}|{\bm U}_{[K/2,4K]})+\mathrm{c.c.}\right],
\label{scale-nonlocal information flow_reversible current expression_IR}
\end{align}
By noting that ${\bm U}_{[K/2,4K]}={\bm U}_{[K/2,K]}\cup{\bm U}_{[2K,4K]}$, we obtain
\begin{align}
\dfrac{p_t({\bm U}^<_{K/4},{\bm U}_{[2K,4K]}|{\bm U}_{[K/2,K]})}{p_t({\bm U}_{[2K,4K]}|{\bm U}_{[K/2,K]})}=p_t({\bm U}^<_{K/4}|{\bm U}_{[K/2,4K]}).
\label{conditional probability relation_2}
\end{align}
By substituting Eq.~(\ref{conditional probability relation_2}) into Eq.~(\ref{scale-nonlocal information flow_reversible current expression_IR}), we obtain
\begin{align}
&\quad\dot{I}_{[2K,4K]}[{\bm U}^<_{K/4}\colon\!{\bm U}_{[2K,4K]}|{\bm U}_{[K/2,K]}]\notag\\
% &=\sum^{n_K+2}_{n=n_K+1}\int dudu^*\left[B_n(u,u^*)p_t(u,u^*)\dfrac{\partial}{\partial u_n}\ln\dfrac{p_t({\bm U}^<_{K/4},{\bm U}_{[2K,4K]}|{\bm U}_{[K/2,K]})}{p_t({\bm U}^<_{K/4}|{\bm U}_{[K/2,K]})p_t({\bm U}_{[2K,4K]}|{\bm U}_{[K/2,K]})}+\mathrm{c.c.}\right]\notag\\
&=\sum^{n_K+2}_{n=n_K+1}\int dudu^*\left[B_n(u,u^*)p_t(u,u^*)\dfrac{\partial}{\partial u_n}\ln\dfrac{p_t({\bm U}^<_{K/4}|{\bm U}_{[K/2,4K]})}{p_t({\bm U}^<_{K/4}|{\bm U}_{[K/2,K]})}+\mathrm{c.c.}\right]\notag\\
&=\sum^{n_K+2}_{n=n_K+1}\int dudu^*\left[B_n(u,u^*)p_t({\bm U}^>_{4K}|{\bm U}^<_{4K})p_t({\bm U}_{[K/2,4K]})\dfrac{\partial}{\partial u_n}p_t({\bm U}^<_{K/4}|{\bm U}_{[K/2,4K]})+\mathrm{c.c.}\right]\notag\\
&=\sum^{n_K+2}_{n=n_K+1}\int d{\bm U}^<_{4K}\left[\overline{B}_n({\bm U}^<_{4K})p_t({\bm U}_{[K/2,4K]})\dfrac{\partial}{\partial u_n}p_t({\bm U}^<_{K/4}|{\bm U}_{[K/2,4K]})+\mathrm{c.c.}\right],
\label{indirect interaction part}
\end{align}
where we used the fact that $\frac{\partial}{\partial u_n}p_t({\bm U}^<_{K/4}|{\bm U}_{[K/2,K]})=0$ for $n=n_K+1,n_K+2$, and introduced the effective nonlinear term $\overline{B}_n({\bm U}^<_{4K})$ as the conditional average of $B_n(u,u^*)$ with respect to the conditional probability density $p_t({\bm U}^>_{4K}|{\bm U}^<_{4K})$:
\begin{align}
\overline{B}_n({\bm U}^<_{4K})=\int d{\bm U}^>_{4K}B_n(u,u^*)p_t({\bm U}^>_{4K}|{\bm U}^<_{4K}).
\label{effective nonlinear term}
\end{align}
\begin{figure}[t]
\centering
\includegraphics[width=10cm]{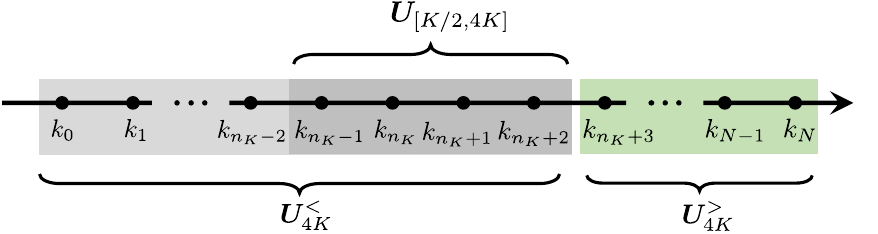}
\caption{Schematic of the approximation of the conditional probability density: $p_t({\bm U}^>_{4K}|{\bm U}^<_{4K})\simeq p_t({\bm U}^>_{4K}|{\bm U}_{[K/2,4K]})$.}
\label{fig:schematic_conditional_probability}
\end{figure}

Note that $\overline{B}_n({\bm U}^<_{4K})$ depends on $u_m$ even if $m\ll n$, and thus includes scale-nonlocal interactions.
However, by assuming Kolmogorov's hypothesis for the Kolmogorov multiplier~\cite{kolmogorov1962refinement}, we can ignore the contribution from the scale-nonlocal interactions.
In the shell model, the multipliers $z_n\in\mathbb{C}$ are defined by $z_n:=|u_n/u_{n-1}|e^{i\Delta_n}$, where $\Delta_n:=\theta_n-\theta_{n-1}-\theta_{n-2}$ with $\theta_n:=\arg u_n$~\cite{benzi1993intermittency,eyink2003gibbsian,biferale2017optimal}.
There is a one-to-one correspondence between the multipliers $\{z_n\}$ and $\{u_n\}$.
% Note that $z_0$ is not defined because $u_{-1}=0$.
Then, Kolmogorov's hypothesis states that the single-time statistics of multipliers is universal and independent of the shell number $n$ in the inertial range and that the multipliers for widely separated shells are statistically independent.
This hypothesis agrees well with numerical simulations~\cite{eyink2003gibbsian,biferale2017optimal}.
Under this assumption, the conditional probability density $p_t({\bm U}^>_{4K}|{\bm U}^<_{4K})$ can be expressed as
\begin{align}
p_t({\bm U}^>_{4K}|{\bm U}^<_{4K})d{\bm U}^>_{4K}=p_{\mathrm{uni}}({\bm Z}^>_{4K}|{\bm Z}^{<,\mathrm{local}}_{4K})d{\bm Z}^>_{4K},
\label{Kolmogorov hypothesis}
\end{align}
where ${\bm Z}^>_{4K}:=\{z_n,z^*_n\mid n_K+3\le n\le N\}$ denotes the small-scale multipliers, and $p_{\mathrm{uni}}({\bm Z}^>_{4K}|{\bm Z}^{<,\mathrm{local}}_{4K})$ denotes the conditional probability density for ${\bm Z}^>_{4K}$, which is universal and time-independent in the developed turbulent regime~\cite{biferale2017optimal}.
Here, we used the notation ${\bm Z}^{<,\mathrm{local}}_{4K}:=\{z_{n_K+2},z^*_{n_K+2},z_{n_K+1},z^*_{n_K+1},\ldots\}$ to indicate that $p_{\mathrm{uni}}({\bm Z}^>_{4K}|{\bm Z}^{<,\mathrm{local}}_{4K})$ has a weak dependence on $z_n$ for $n\ll n_K+2$~\cite{biferale2017optimal}.
By substituting Eq.~(\ref{Kolmogorov hypothesis}) into Eq.~(\ref{effective nonlinear term}), we find that $\overline{B}_n({\bm U}^<_{4K})$ depends essentially on a few variables $u_{n_K+2},u^*_{n_K+2},u_{n_K+1},u^*_{n_K+1},\ldots$.
Therefore, $\overline{B}_n({\bm U}^<_{4K})$ can be approximated by $\overline{B}_n({\bm U}_{[K/2,4K]})$, which is equivalent to $p_t({\bm U}^>_{4K}|{\bm U}^<_{4K})\simeq p_t({\bm U}^>_{4K}|{\bm U}_{[K/2,4K]})$ (see Fig.~\ref{fig:schematic_conditional_probability}).
Then, Eq.~(\ref{indirect interaction part}) becomes
% According to the numerical study in Ref.~\cite{biferale2017optimal}, the conditional probability density $p_t({\bm U}^>_{4K}|{\bm U}^<_{4K})$ can be approximated by $p_t({\bm U}^>_{4K}|{\bm U}_{[K/2,4K]})$, i.e., $p_t({\bm U}^>_{4K}|{\bm U}^<_{4K})$ has a short-range dependence on the large-scale modes ${\bm U}^<_{4K}$ (see Fig.~\ref{fig:schematic_conditional_probability}).
% Then, we have $\overline{B}_n({\bm U}^<_{4K})\simeq\overline{B}_n({\bm U}_{[K/2,4K]})$, and Eq.~(\ref{indirect interaction part}) becomes
\begin{align}
&\quad\dot{I}_{[2K,4K]}[{\bm U}^<_{K/4}\colon\!{\bm U}_{[2K,4K]}|{\bm U}_{[K/2,K]}]\notag\\
&\simeq\sum^{n_K+2}_{n=n_K+1}\int d{\bm U}^<_{K/4}d{\bm U}_{[K/2,4K]}\left[\overline{B}_n({\bm U}_{[K/2,4K]})p_t({\bm U}_{[K/2,4K]})\dfrac{\partial}{\partial u_n}p_t({\bm U}^<_{K/4}|{\bm U}_{[K/2,4K]})+\mathrm{c.c.}\right]\notag\\
&=\sum^{n_K+2}_{n=n_K+1}\int d{\bm U}_{[K/2,4K]}\left[\overline{B}_n({\bm U}_{[K/2,4K]})p_t({\bm U}_{[K/2,4K]})\dfrac{\partial}{\partial u_n}\int d{\bm U}^<_{K/4}p_t({\bm U}^<_{K/4}|{\bm U}_{[K/2,4K]})+\mathrm{c.c.}\right]\notag\\
&=0,
\end{align}
where we have used $\frac{\partial}{\partial u_n}\int d{\bm U}^<_{K/4}p_t({\bm U}^<_{K/4}|{\bm U}_{[K/2,4K]})=\frac{\partial}{\partial u_n}1=0$.
Therefore, the scale-nonlocal information flow can be ignored, and we finally arrive at the main result~(\ref{scale locality}).

\section{Concluding remarks\label{Concluding remarks}}
% summary
In this study, we proved the scale locality of the scale-to-scale information flow for the shell model.
Specifically, we showed that the scale-to-scale information flow can be decomposed into scale-local and scale-nonlocal parts and that the scale-nonlocal contribution can be ignored compared to the scale-local part under the Kolmogorov hypothesis.
This result suggests that turbulent fluctuations at large scales do not directly influence those at small scales without scale-local interactions.
Although it remains unclear whether the scale locality is directly involved in the generating mechanism of the universal statistics, our result can be regarded at least as a first step toward understanding how universal scaling laws of turbulent fluctuations emerge at small scales under the influence of the information flow from large scales.
% We believe that this result is essential for understanding how universal scaling laws of turbulent fluctuations emerge at small scales under the influence of the information flow from large scales. 
% For example, it seems reasonable to conjecture that the universal scaling laws result from the accumulation of errors during the scale-local transmission process.

% We now provide several remarks on our results.
Because our proof is based on the scale locality of the nonlinear interaction, it cannot be directly applied to the case of the Navier--Stokes equation, where the nonlinear interaction is scale-nonlocal.
Even in this case, by identifying the scale-local modes as some band-pass filtered Fourier modes, we can formally decompose the scale-to-scale information flow into scale-local and scale-nonlocal parts as in Sec.~\ref{subsec: Decomposition of the information flow into scale-local and scale-nonlocal parts}.
Then, because the Kolmogorov hypothesis agrees well with experimental and direct numerical simulation results~\cite{chen2003kolmogorov,benzi1998multiscale,pedrizzetti1996self,chhabra1992scale}, it may be possible to clarify some aspect of the scale locality of the information flow in the inertial range.
% Nevertheless, it seems reasonable to conjecture that the scale locality of the information flow is satisfied even in this case because the Kolmogorov hypothesis agrees well with experimental and direct numerical simulation results~\cite{chen2003kolmogorov,benzi1998multiscale,pedrizzetti1996self,chhabra1992scale}.
The scale locality of the energy cascade also supports this conjecture~\cite{domaradzki1990local,eyink2005locality,eyink2009localness,aluie2009localness,cardesa2017turbulent,goto2017hierarchy,johnson2020energy,johnson2021role}.

The relation between the scale locality of the information transfer and that of the energy cascade, or more generally, the scale locality of the cascade of inviscid conserved quantities, remains an open problem.
It would be an interesting research direction to investigate whether a situation in which the cascade is scale local but the information transfer is scale nonlocal, or vice versa, can be realized. 
In this regard, it is desirable to elucidate how information flows in the case of other types of turbulence, such as two-dimensional turbulence~\cite{tabeling2002two,boffetta2012two,alexakis2018cascades} and quantum turbulence~\cite{tsubota2013quantum,QT_review,Tsatsos_2016,skrbek2021phenomenology}.
We also note that it is possible to investigate the scale locality of various types of cascades by considering the mutual information or information flow for the corresponding fluxes.
Indeed, several previous studies have numerically illustrated that (generalized) transfer entropy for the energy flux can capture the locality of the energy cascade more accurately than time cross-correlation functions~\cite{PhysRevResearch.4.023195,araki2024forgetfulness}.

% \section*{Acknowledgments}
\ack
T.T.~thanks Ryo Araki for fruitful discussions.
T.T.~was supported by JSPS KAKENHI Grant Number JP23K19035, JP25K17315, and JST PRESTO Grant Number JPMJPR23O6, Japan.

% \newpage
\appendix

\section{Expression for information flow in terms of probability currents\label{Expression for information flow in terms of probability currents}}
In this section, we derive an expression for information flow in terms of probability currents for general Langevin equations in the following form:
\begin{align}
\partial_tx_t&=F^X_t(x_t,y_t)+\sqrt{2D^X}\xi^X_t,\\
\partial_ty_t&=F^Y_t(x_t,y_t)+\sqrt{2D^Y}\xi^Y_t,
\end{align}
where $\xi^\alpha$ ($\alpha=X,Y$) denotes a zero-mean white Gaussian noise that satisfies $\langle\xi^{\alpha}_t\xi^{\alpha'}_{t'}\rangle=\delta_{\alpha\alpha'}\delta(t-t')$.
Here, we have assumed that $x, y\in \mathbb{R}$ for simplicity.
The extension to complex multidimensional variables such as ${\bm U}^<_K$ and ${\bm U}^>_K$ in the shell model is straightforward (see Appendix B of Ref.~\cite{tanogami2024information}).
The corresponding Fokker--Planck equation reads
\begin{align}
\partial_tp_t(x,y)=-\partial_xJ^X_t(x,y)-\partial_yJ^Y_t(x,y),
\end{align}
where $J^\alpha_t(x,y)$ denotes the probability current associated with $\alpha$ ($\partial_\alpha$ denotes $\partial_x$ or $\partial_y$),
\begin{align}
J^\alpha_t(x,y)=F^\alpha_t(x,y)p_t(x,y)-D^\alpha\partial_\alpha p_t(x,y).
\end{align}

Let $\dot{I}^X[X\colon\!Y]$ ($\dot{I}^Y[X\colon\!Y]$) be the information flow from $Y$ to $X$ ($X$ to $Y$) defined by
\begin{align}
\dot{I}^X[X\colon\!Y]&:=\lim_{\Delta t\rightarrow0}\dfrac{I[X_{t+\Delta t}\colon\!Y_t]-I[X_t\colon\!Y_t]}{\Delta t},\label{IF definition X}\\
\dot{I}^Y[X\colon\!Y]&:=\lim_{\Delta t\rightarrow0}\dfrac{I[X_t\colon\!Y_{t+\Delta t}]-I[X_t\colon\!Y_t]}{\Delta t}.\label{IF definition Y}
\end{align}
We now show that $\dot{I}^X[X\colon\!Y]$ and $\dot{I}^Y[X\colon\!Y]$ can be expressed as
\begin{align}
\dot{I}^X[X\colon\!Y]&=\int dxdyJ^X_t(x,y)\partial_x\ln\dfrac{p_t(x,y)}{p^X_t(x)p^Y_t(y)},\label{IF expression X}\\
\dot{I}^Y[X\colon\!Y]&=\int dxdyJ^Y_t(x,y)\partial_y\ln\dfrac{p_t(x,y)}{p^X_t(x)p^Y_t(y)},\label{IF expression Y}
\end{align}
where $p^X_t(x)=\int dyp_t(x,y)$ and $p^Y_t(y)=\int dxp_t(x,y)$ are marginal distributions.
From the definition (\ref{IF definition X}), we first note that
\begin{align}
\dot{I}^X[X\colon\!Y]=\lim_{h\rightarrow0}\dfrac{1}{h}&\left(\int dxdy'p(x,t+h;y',t)\ln\dfrac{p(x,t+h;y',t)}{p^X_{t+h}(x)p^Y_t(y')}\right.\notag\\
&\qquad\left.-\int dx'dy'p_t(x',y')\ln\dfrac{p_t(x',y')}{p^X_t(x')p^Y_t(y')}\right),
\end{align}
where $p(x,t+h;y',t)$ denotes the two-point probability density.
When $h=0$, the two-point probability density corresponds to the joint probability density at time $t$: $p(x,t;y',t)=p_t(x,y')$.
By expanding $p(x,t+h;y',t)$ and $p^X_{t+h}(x)$ with respect to $h$, we obtain
\begin{align}
\dot{I}^X[X\colon\!Y]&=\int dxdy'\left.\dfrac{d}{dh}p(x,t+h;y',t)\right|_{h=0}\ln\dfrac{p_t(x,y')}{p^X_t(x)p^Y_t(y')}\notag\\
&\qquad+\int dxdy'\left.\dfrac{d}{dh}p(x,t+h;y',t)\right|_{h=0}-\int dxdy'\dfrac{p_t(x,y')}{p^X_t(x)}\left.\dfrac{d}{dh}p^X_{t+h}(x)\right|_{h=0}.
\label{IF expanding in h}
\end{align}
The second and third terms on the right-hand side of Eq.~(\ref{IF expanding in h}) vanish because $\frac{d}{dh}\int dxdy'p(x,t+h;y',t)=\frac{d}{dh}\int dxp^X_{t+h}(x)=\frac{d}{dh}1=0$.
By noting that
\begin{align}
p(x,t+h;y',t)=\int dydx'p(x,y,t+h;x',y',t)=\int dydx'p(x,y,t+h|x',y',t)p_t(x',y'),
\end{align}
and that the conditional probability density $p(x,y,t+h|x',y',t)$ obeys the Fokker--Planck equation, we obtain
\begin{align}
\dot{I}^X[X\colon\!Y]&=\int dxdydx'dy'\left.\dfrac{d}{dh}p(x,y,t+h|x',y',t)\right|_{h=0}p_t(x',y')\ln\dfrac{p_t(x,y')}{p^X_t(x)p^Y_t(y')}\notag\\
&=-\int dxdydx'dy'\partial_x\left[F^X_t(x,y)\delta(x-x')\delta(y-y')-D^X\partial_x\delta(x-x')\delta(y-y')\right]\notag\\
&\qquad\times p_t(x',y')\ln\dfrac{p_t(x,y')}{p^X_t(x)p^Y_t(y')}\notag\\
&\quad-\int dxdydx'dy'\partial_y\left[F^Y_t(x,y)\delta(x-x')\delta(y-y')-D^Y\partial_y\delta(x-x')\delta(y-y')\right]\notag\\
&\qquad\times p_t(x',y')\ln\dfrac{p_t(x,y')}{p^X_t(x)p^Y_t(y')}\notag\\
&=-\int dxdy\partial_x\left[F^X_t(x,y)p_t(x,y)-D^X\partial_xp_t(x,y)\right]\ln\dfrac{p_t(x,y)}{p^X_t(x)p^Y_t(y)}\notag\\
&\quad-\int dxdy\partial_y\left[F^Y_t(x,y)p_t(x,y)\ln\dfrac{p_t(x,y)}{p^X_t(x)p^Y_t(y)}-D^Y\partial_y\left(p_t(x,y)\ln\dfrac{p_t(x,y)}{p^X_t(x)p^Y_t(y)}\right)\right]\notag\\
&=\int dxdyJ^X_t(x,y)\partial_x\ln\dfrac{p_t(x,y)}{p^X_t(x)p^Y_t(y)},
\end{align}
where we used the fact that the second term on the right-hand side of the third equality vanishes, provided that the terms in the square bracket vanish at infinity.
We thus arrive at Eq.~(\ref{IF expression X}) (Eq.~(\ref{IF expression Y}) can be derived in the same way).
We can also show the expression for the conditional information flow [Eqs.~(\ref{conditional IF expression J_X}) and (\ref{conditional IF expression J_Y})] following the same argument.

We remark that Eqs.~(\ref{IF expression X}) and (\ref{IF expression Y}) can also be quickly derived by noting that the time derivative of the \textit{stochastic mutual information} $\hat{I}_t(x_t\colon\!y_t):=\ln(p_t(x_t,y_t)/p^X_t(x_t)p^Y_t(y_t))$ reads
\begin{align}
\dfrac{d}{dt}\hat{I}_t(x_t\colon\!y_t)=\hat{\dot{I}}^X_t+\hat{\dot{I}}^Y_t,
\label{stochastic mutual information_derivative}
\end{align}
where $\hat{\dot{I}}^X_t$ and $\hat{\dot{I}}^Y_t$ denote the \textit{stochastic information flow} defined by
\begin{align}
\hat{\dot{I}}^X_t&=\dot{x}_t\circ\partial_x\ln\dfrac{p_t(x_t,y_t)}{p^X_t(x_t)p^Y_t(y_t)}-\dfrac{1}{p_t(x_t,y_t)}\partial_xJ^X_t(x_t,y_t)+\dfrac{1}{p^X_t(x_t)}\partial_x\bar{J}^X_t(x_t),\\
\hat{\dot{I}}^Y_t&=\dot{y}_t\circ\partial_y\ln\dfrac{p_t(x_t,y_t)}{p^X_t(x_t)p^Y_t(y_t)}-\dfrac{1}{p_t(x_t,y_t)}\partial_yJ^Y_t(x_t,y_t)+\dfrac{1}{p^Y_t(y_t)}\partial_y\bar{J}^Y_t(y_t).
\end{align}
Here, the symbol $\circ$ denotes multiplication in the sense of Stratonovich~\cite{gardiner1985handbook}, and $\bar{J}^X_t(x):=\int dyJ^X_t(x,y)$ and $\bar{J}^Y_t(y):=\int dxJ^Y_t(x,y)$ denote the effective probability currents.
The (mean) information flow can be obtained by taking the average of the stochastic information flow.
By noting that $\langle\dot{x}_t\circ g_t(x_t,y_t)\rangle=\int dxdyJ^X_t(x,y)g_t(x,y)$ for any function $g_t(x,y)$~\cite{seifert2012stochastic}, we arrive at Eqs.~(\ref{IF expression X}) and (\ref{IF expression Y}).

\section{Detailed derivation of Eq.~(\ref{IF=IF_rev})\label{Detailed discussion on irreversible information flow}}
In this section, we provide a detailed derivation of Eq.~(\ref{IF=IF_rev}).
To this end, we first nondimensionalize the shell model with large-scale characteristic quantities $k_f$ and $U:=\varepsilon^{1/3}k^{-1/3}_f$ by setting
\begin{align}
\hat{u}_n:=u_n/U,\quad \hat{k}_n:=k_n/k_f,\quad \hat{t}:=t/\tau_L,\quad\hat{\xi}_n:=\tau^{1/2}_L\xi_n,\quad \hat{f}_n:=f_n/(U^2k_f),
\end{align}
where $\tau_L:=1/k_fU$ denotes the large-eddy turnover time.
The nondimensionalized Eq.~(\ref{sabra shell model}) reads
\begin{align}
\partial_{\hat{t}}{\hat{u}}_n&=\hat{B}_n(\hat{u},\hat{u}^*)-\dfrac{1}{\mathrm{Re}}\hat{k}^2_n\hat{u}_n+\sqrt{2\Theta}\hat{k}_n\hat{\xi}_n+\hat{f}_n,
\label{sabra shell model_dimensionless}
\end{align}
where
\begin{align}
\hat{B}_n(\hat{u},\hat{u}^*)&:=i\biggl(\hat{k}_{n+1}\hat{u}_{n+2}\hat{u}^*_{n+1}-\dfrac{1}{2}\hat{k}_n\hat{u}_{n+1}\hat{u}^*_{n-1}+\dfrac{1}{2}\hat{k}_{n-1}\hat{u}_{n-1}\hat{u}_{n-2}\biggr).
\end{align}
Here, $\mathrm{Re}:=U/(\nu k_f)$ denotes the Reynolds number, and $\Theta:=\mathrm{Re}^{-3/2}\theta_\nu$ denotes the dimensionless temperature, where $\theta_\nu:=k_{\mathrm{B}}T/\rho(\varepsilon\nu)^{1/2}$ denotes the ratio of the thermal energy to the kinetic energy at the energy dissipation scale $k_\nu$.
Note that a typical value of $\theta_\nu$ for the turbulent atmospheric boundary layer is on the order of $10^{-8}$~\cite{bandak2022dissipation}. 
The corresponding Fokker--Planck equation reads
\begin{align}
\partial_{\hat{t}}p_{\hat{t}}(\hat{u},\hat{u}^*)=\sum^N_{n=0}\left[-\dfrac{\partial}{\partial \hat{u}^*_n}\hat{J}_n(\hat{u},\hat{u}^*)-\dfrac{\partial}{\partial \hat{u}^*_n}\hat{J}^*_n(\hat{u},\hat{u}^*)\right],
\end{align}
where $\hat{J}_n(\hat{u},\hat{u}^*)$ denotes the dimensionless probability current associated with $\hat{u}_n$:
\begin{align}
\hat{J}_n(\hat{u},\hat{u}^*)&:=\left(\hat{B}_n(\hat{u},\hat{u}^*)-\dfrac{1}{\mathrm{Re}}\hat{k}^2_n\hat{u}_n+\hat{f}_n\right)p_{\hat{t}}(\hat{u},\hat{u}^*)-2\Theta \hat{k}^2_n\dfrac{\partial}{\partial \hat{u}^*_n}p_{\hat{t}}(\hat{u},\hat{u}^*)\notag\\
&=\hat{J}^{\mathrm{rev}}_n(\hat{u},\hat{u}^*)+\hat{J}^{\mathrm{irr}}_n(\hat{u},\hat{u}^*).
\end{align}
Here, $\hat{J}^{\mathrm{rev}}_n(\hat{u},\hat{u}^*)$ and $\hat{J}^{\mathrm{irr}}_n(\hat{u},\hat{u}^*)$ denote the dimensionless reversible and irreversible currents, respectively:
\begin{align}
\hat{J}^{\mathrm{rev}}_n(\hat{u},\hat{u}^*)&=\left(\hat{B}_n(\hat{u},\hat{u}^*)+\hat{f}_n\right)p_{\hat{t}}(\hat{u},\hat{u}^*),\\
\hat{J}^{\mathrm{irr}}_n(\hat{u},\hat{u}^*)&=-\dfrac{1}{\mathrm{Re}}\hat{k}^2_n\hat{u}_np_{\hat{t}}(\hat{u},\hat{u}^*)-2\Theta \hat{k}^2_n\dfrac{\partial}{\partial \hat{u}^*_n}p_{\hat{t}}(\hat{u},\hat{u}^*).
\label{dimensionless J_irr}
\end{align}
Then, the dimensionless reversible or irreversible part of the information flow is given by
\begin{align}
\hat{\dot{I}}^{<,\mathrm{rev/irr}}_K&=\sum^{n_K}_{n=0}\int d\hat{u}d\hat{u}^*\left[\hat{J}^{\mathrm{rev/irr}}_n(\hat{u},\hat{u}^*)\dfrac{\partial}{\partial \hat{u}_n}\ln\dfrac{p_{\hat{t}}(\hat{u},\hat{u}^*)}{p_{\hat{t}}(\hat{{\bm U}}^<_K)p_{\hat{t}}(\hat{{\bm U}}^>_K)}+\mathrm{c.c.}\right].
% \hat{\dot{I}}^{<,\mathrm{rev/irr}}_K&=\sum^{n_K}_{n=0}\int d\hat{u}d\hat{u}^*\left[\hat{J}^{\mathrm{rev/irr}}_n(\hat{u},\hat{u}^*)\dfrac{\partial}{\partial \hat{u}_n}\ln\dfrac{p_{\hat{t}}(\hat{u},\hat{u}^*)}{p_{\hat{t}}(\hat{{\bm U}}^<_K)p_{\hat{t}}(\hat{{\bm U}}^>_K)}+\hat{J}^{\mathrm{rev/irr}*}_n(\hat{u},\hat{u}^*)\dfrac{\partial}{\partial \hat{u}^*_n}\ln\dfrac{p_{\hat{t}}(\hat{u},\hat{u}^*)}{p_{\hat{t}}(\hat{{\bm U}}^<_K)p_{\hat{t}}(\hat{{\bm U}}^>_K)}\right].
\end{align}
We consider the limit $\mathrm{Re}\rightarrow\infty$ with $K/k_f$ and $\theta_\nu$ held fixed~\cite{bandak2024spontaneous}.
Then, from the expression of Eq.~(\ref{dimensionless J_irr}), we find that $\hat{\dot{I}}^{<,\mathrm{irr}}_K\rightarrow0$.

\section*{References}
\bibliographystyle{iopart-num}
\bibliography{main_text}

\providecommand{\newblock}{}
\begin{thebibliography}{10}
\expandafter\ifx\csname url\endcsname\relax
  \def\url#1{{\tt #1}}\fi
\expandafter\ifx\csname urlprefix\endcsname\relax\def\urlprefix{URL }\fi
\providecommand{\eprint}[2][]{\url{#2}}

\bibitem{Frisch}
Frisch U 1995 {\em Turbulence\/} (Cambridge university press)

\bibitem{davidson2015turbulence}
Davidson P~A 2015 {\em Turbulence: {A}n {I}ntroduction for {S}cientists and
  {E}ngineers\/} 2nd ed (Oxford University Press)

\bibitem{Eyink_lecture}
Eyink G~L Turbulence {T}heory, {C}ourse {N}otes
  \url{http://www.ams.jhu.edu/~eyink/Turbulence/notes/}

\bibitem{Eyink_Sreenivasan}
Eyink G~L and Sreenivasan K~R 2006 {\em Rev. Mod. Phys.\/} {\bf 78} 87

\bibitem{tanogami2024information}
Tanogami T and Araki R 2024 {\em Phys. Rev. Research\/} {\bf 6} 013090

\bibitem{tanogami2025amplify}
Tanogami T and Araki R 2025 {\em Phys. Rev. Research\/} {\bf 7}(2) 023078
  \urlprefix\url{https://link.aps.org/doi/10.1103/PhysRevResearch.7.023078}

\bibitem{cover1999elements}
Cover T~M and Thomas J~A 2006 {\em Elements of {I}nformation {T}heory\/} 2nd ed
  (Wiley-Interscience, Hoboken, NJ)

\bibitem{peliti2021stochastic}
Peliti L and Pigolotti S 2021 {\em Stochastic {T}hermodynamics: {A}n
  {I}ntroduction\/} (Princeton University Press)

\bibitem{shiraishi2023introduction}
Shiraishi N 2023 {\em {An Introduction to Stochastic Thermodynamics: From Basic
  to Advanced}\/} vol 212 (Springer Nature)

\bibitem{parrondo2015thermodynamics}
Parrondo J~M, Horowitz J~M and Sagawa T 2015 {\em Nat. Phys.\/} {\bf 11}
  131--139

\bibitem{horowitz2014thermodynamics}
Horowitz J~M and Esposito M 2014 {\em Phys. Rev. X\/} {\bf 4} 031015

\bibitem{ehrich2023energy}
Ehrich J and Sivak D~A 2023 {\em Front. Phys.\/} {\bf 11} 155

\bibitem{domaradzki1990local}
Domaradzki J~A and Rogallo R~S 1990 {\em Phys. Fluids A\/} {\bf 2} 413--426

\bibitem{eyink2005locality}
Eyink G~L 2005 {\em Physica D\/} {\bf 207} 91--116

\bibitem{eyink2009localness}
Eyink G~L and Aluie H 2009 {\em Phys. Fluids\/} {\bf 21}

\bibitem{aluie2009localness}
Aluie H and Eyink G~L 2009 {\em Phys. Fluids\/} {\bf 21}

\bibitem{cardesa2017turbulent}
Cardesa J~I, Vela-Mart{\'\i}n A and Jim{\'e}nez J 2017 {\em Science\/} {\bf
  357} 782--784

\bibitem{goto2017hierarchy}
Goto S, Saito Y and Kawahara G 2017 {\em Phys. Rev. Fluids\/} {\bf 2} 064603

\bibitem{johnson2020energy}
Johnson P~L 2020 {\em Phys. Rev. Lett.\/} {\bf 124} 104501

\bibitem{johnson2021role}
Johnson P~L 2021 {\em J. Fluid Mech.\/} {\bf 922} A3

\bibitem{kolmogorov1962refinement}
Kolmogorov A~N 1962 {\em J. Fluid Mech.\/} {\bf 13} 82--85

\bibitem{chen2003kolmogorov}
Chen Q, Chen S, Eyink G~L and Sreenivasan K~R 2003 {\em Phys. Rev. Lett.\/}
  {\bf 90} 254501

\bibitem{benzi1998multiscale}
Benzi R, Biferale L and Toschi F 1998 {\em Phys. Rev. Lett.\/} {\bf 80} 3244

\bibitem{pedrizzetti1996self}
Pedrizzetti G, Novikov E~A and Praskovsky A~A 1996 {\em Phys. Rev. E\/} {\bf
  53} 475

\bibitem{chhabra1992scale}
Chhabra A~B and Sreenivasan K 1992 {\em Phys. Rev. Lett.\/} {\bf 68} 2762

\bibitem{bandak2022dissipation}
Bandak D, Goldenfeld N, Mailybaev A~A and Eyink G 2022 {\em Phys. Rev. E\/}
  {\bf 105} 065113

\bibitem{bell2022thermal}
Bell J~B, Nonaka A, Garcia A~L and Eyink G 2022 {\em J. Fluid Mech.\/} {\bf
  939}

\bibitem{mcmullen2022navier}
McMullen R~M, Krygier M~C, Torczynski J~R and Gallis M~A 2022 {\em Phys. Rev.
  Lett.\/} {\bf 128} 114501

\bibitem{gledzer1973system}
Gledzer E~B 1973 System of hydrodynamic type admitting two quadratic integrals
  of motion {\em Sov. Phys. Doklady\/} vol~18 p 216

\bibitem{ohkitani1989temporal}
Ohkitani K and Yamada M 1989 {\em Prog. Theor. Phys.\/} {\bf 81} 329--341

\bibitem{komatsu2014glimpse}
Komatsu T~S, Matsumoto S, Shimada T and Ito N 2014 {\em Int. J. Mod. Phys. C\/}
  {\bf 25} 1450034

\bibitem{ruelle1979microscopic}
Ruelle D 1979 {\em Phys. Lett. A\/} {\bf 72} 81--82

\bibitem{lorenz1969predictability}
Lorenz E~N 1969 {\em Tellus\/} {\bf 21} 289--307

\bibitem{bandak2024spontaneous}
Bandak D, Mailybaev A~A, Eyink G~L and Goldenfeld N 2024 {\em Phys. Rev.
  Lett.\/} {\bf 132} 104002

\bibitem{palmer2024real}
Palmer T 2024 {\em Phys. Today\/} {\bf 77} 30--35

\bibitem{Srivastava2025_molecular}
Srivastava I, Nonaka A~J, Zhang W, Garcia A~L and Bell J~B 2025 {\em arXiv
  preprint arXiv:2501.06396\/}

\bibitem{l1998improved}
L'vov V~S, Podivilov E, Pomyalov A, Procaccia I and Vandembroucq D 1998 {\em
  Phys. Rev. E\/} {\bf 58} 1811

\bibitem{bandak2021thermal}
Bandak D, Eyink G~L, Mailybaev A and Goldenfeld N 2021 {\em arXiv preprint
  arXiv:2107.03184\/}

\bibitem{maes2021local}
Maes C 2021 {\em SciPost Phys. Lect. Notes\/} {\bf 32} 1

\bibitem{landau1959fluid}
Landau L~D and Lifshitz E~M 1959 {\em Fluid {M}echanics\/} vol~6
  (Addision-Wesley, Reading, MA)

\bibitem{de2006hydrodynamic}
De~Zarate J~M~O and Sengers J~V 2006 {\em Hydrodynamic fluctuations in fluids
  and fluid mixtures\/} (Elsevier)

\bibitem{risken1996fokker}
Risken H 1996 The {F}okker-{P}lanck {E}quation (Springer)

\bibitem{bohr1998dynamical}
Bohr T, Jensen M~H, Paladin G and Vulpiani A 1998 {\em Dynamical systems
  approach to turbulence\/} (Cambridge university press)

\bibitem{biferale2003shell}
Biferale L 2003 {\em Annu. Rev. Fluid Mech.\/} {\bf 35} 441--468

\bibitem{schreiber2000measuring}
Schreiber T 2000 {\em Phys. Rev. Lett.\/} {\bf 85} 461

\bibitem{materassi2014information}
Materassi M, Consolini G, Smith N and De~Marco R 2014 {\em Entropy\/} {\bf 16}
  1272--1286

\bibitem{PhysRevResearch.4.023195}
Lozano-Dur\'an A and Arranz G 2022 {\em Phys. Rev. Research\/} {\bf 4}(2)
  023195
  \urlprefix\url{https://link.aps.org/doi/10.1103/PhysRevResearch.4.023195}

\bibitem{araki2024forgetfulness}
Araki R, Vela-Mart{\'\i}n A and Lozano-Dur{\'a}n A 2024 Forgetfulness of
  turbulent energy cascade associated with different mechanisms {\em J. Phys.
  Conf.\/} vol 2753 (IOP Publishing) p 012001

\bibitem{chetrite2019information}
Chetrite R, Rosinberg M~L, Sagawa T and Tarjus G 2019 {\em J. Stat. Mech.\/}
  {\bf 2019} 114002

\bibitem{smirnov2013spurious}
Smirnov D~A 2013 {\em Phys. Rev. E\/} {\bf 87} 042917

\bibitem{james2016information}
James R~G, Barnett N and Crutchfield J~P 2016 {\em Phys. Rev. Lett.\/} {\bf
  116} 238701

\bibitem{allahverdyan2009thermodynamic}
Allahverdyan A~E, Janzing D and Mahler G 2009 {\em J. Stat. Mech.\/} {\bf 2009}
  P09011

\bibitem{hartich2014stochastic}
Hartich D, Barato A~C and Seifert U 2014 {\em J. Stat. Mech.\/} {\bf 2014}
  P02016

\bibitem{barato2014efficiency}
Barato A~C, Hartich D and Seifert U 2014 {\em New J. Phys.\/} {\bf 16} 103024

\bibitem{horowitz2015multipartite}
Horowitz J~M 2015 {\em J. Stat. Mech.\/}  P03006

\bibitem{hartich2016sensory}
Hartich D, Barato A~C and Seifert U 2016 {\em Phys. Rev. E\/} {\bf 93} 022116

\bibitem{matsumoto2018role}
Matsumoto T and Sagawa T 2018 {\em Phys. Rev. E\/} {\bf 97} 042103

\bibitem{benzi1993intermittency}
Benzi R, Biferale L and Parisi G 1993 {\em Physica D\/} {\bf 65} 163--171

\bibitem{eyink2003gibbsian}
Eyink G~L, Chen S and Chen Q 2003 {\em J. Stat. Phys.\/} {\bf 113} 719--740

\bibitem{biferale2017optimal}
Biferale L, Mailybaev A~A and Parisi G 2017 {\em Phy. Rev. E\/} {\bf 95} 043108

\bibitem{tabeling2002two}
Tabeling P 2002 {\em Phys. Rep.\/} {\bf 362} 1--62

\bibitem{boffetta2012two}
Boffetta G and Ecke R~E 2012 {\em Annu. Rev. Fluid Mech.\/} {\bf 44} 427--451

\bibitem{alexakis2018cascades}
Alexakis A and Biferale L 2018 {\em Phys. Rep.\/} {\bf 767} 1--101

\bibitem{tsubota2013quantum}
Tsubota M, Kobayashi M and Takeuchi H 2013 {\em Phys. Rep.\/} {\bf 522}
  191--238

\bibitem{QT_review}
Barenghi C~F, Skrbek L and Sreenivasan K~R 2014 {\em Proc. Natl. Acad. Sci. U.
  S. A.\/} {\bf 111} 4647--4652

\bibitem{Tsatsos_2016}
Tsatsos M~C, Tavares P~E~S, Cidrim A, Fritsch A~R, Caracanhas M~A, dos Santos
  F~E~A, Barenghi C~F and Bagnato V~S 2016 {\em Phys. Rep.\/} {\bf 622} 1--52

\bibitem{skrbek2021phenomenology}
Skrbek L, Schmoranzer D, Midlik {\v{S}} and Sreenivasan K~R 2021 {\em Proc.
  Natl. Acad. Sci. U. S. A.\/} {\bf 118}

\bibitem{gardiner1985handbook}
Gardiner C~W 2009 {\em Handbook of {S}tochastic {M}ethods\/} 4th ed (Springer,
  Berlin)

\bibitem{seifert2012stochastic}
Seifert U 2012 {\em Rep. Prog. Phys.\/} {\bf 75} 126001

\end{thebibliography}

\end{document}